# The ultra-long GRB 220627A at $z = 3.08$

S. de Wet[1,⋆], L. Izzo[2], P.J. Groot[1,3,4], S. Bisero[5], V. D'Elia[6,7], M. De Pasquale[8], D.H. Hartmann[9], K.E. Heintz[10,11], P. Jakobsson[12], T. Laskar[13], A. Levan[3], A. Martin-Carrillo[14], A. Melandri[7], A. Nicuesa Guelbenzu[15], G. Pugliese[16], A. Rossi[17], A. Saccardi[5], S. Savaglio[17,18,19], P. Schady[20], N.R. Tanvir[21], H. van Eerten[20], and S. Vergani[5]

1. Inter-University Institute for Data Intensive Astronomy & Department of Astronomy, University of Cape Town, Private Bag X3, Rondebosch, 7701, South Africa
2. DARK, Niels Bohr Institute, University of Copenhagen, Jagtvej 128, 2200 Copenhagen, Denmark
3. Department of Astrophysics/IMAPP, Radboud University, P.O. Box 9010, 6500 GL, Nijmegen, The Netherlands
4. South African Astronomical Observatory, P.O. Box 9, 7935, Observatory, South Africa
5. GEPI, Observatoire de Paris, Université PSL, CNRS, 5 Place Jules Janssen, 92190 Meudon, France
6. Space Science Data Center (SSDC) - Agenzia Spaziale Italiana (ASI), 00133 Roma, Italy
7. INAF - Osservatorio Astronomico di Roma, Via Frascati 33, 00040 Monte Porzio Catone, Italy
8. Mathematics, Informatics, Physics, and Earth Science Department, University of Messina, Polo Papardo, Via F. S. D'Alcontres 31, 98166 Messina, Italy
9. Department of Physics and Astronomy, Clemson University, Clemson, SC 29634, USA
10. Cosmic Dawn Center (DAWN), Denmark
11. Niels Bohr Institute, University of Copenhagen, Jagtvej 128, 2200 Copenhagen, Denmark
12. Centre for Astrophysics and Cosmology, Science Institute, University of Iceland, Dunhagi 5, 107 Reykjavík, Iceland
13. Department of Physics & Astronomy, University of Utah, Salt Lake City, UT 84112, USA
14. School of Physics and Centre for Space Research, University College Dublin, Belfield, D04 V1W8 Dublin, Ireland
15. Thüringer Landessternwarte Tautenburg, Sternwarte 5, 07778 Tautenburg, Germany
16. Astronomical Institute Anton Pannekoek, University of Amsterdam, 1090 GE Amsterdam, The Netherlands
17. INAF – Osservatorio di Astrofisica e Scienza dello Spazio, Via Piero Gobetti 93/3, 40129 Bologna, Italy
18. Department of physics, University of Calabria, Via P. Bucci, Arcavacata di Rende (CS), Italy
19. INFN - Laboratori Nazionali di Frascati, Frascati, Italy
20. Physics Department, University of Bath, Claverton Down, Bath, BA2 7AY, UK
21. School of Physics and Astronomy, University of Leicester, University Road, Leicester, LE1 7RH, United Kingdom

**ABSTRACT**

*Context.* GRB 220627A is a rare burst with two distinct γ-ray emission episodes separated by almost 1000 s that triggered the *Fermi* Gamma-ray Burst Monitor twice. High-energy GeV emission was detected by the *Fermi* Large Area Telescope coincident with the first emission episode but not the second. The discovery of the optical afterglow with MeerLICHT led to MUSE observations which secured the burst redshift to $z = 3.08$, making this the most distant ultra-long gamma-ray burst (GRB) detected to date.
*Aims.* The progenitors of some ultra-long GRBs have been suggested in the literature to be different to those of normal long GRBs. Our aim is to determine whether the afterglow and host properties of GRB 220627A agree with this interpretation.
*Methods.* We performed empirical and theoretical modelling of the afterglow data within the external forward shock framework, and determined the metallicity of the GRB environment through modelling the absorption lines in the MUSE spectrum.
*Results.* Our optical data show evidence for a jet break in the light curve at ∼1.2 days, while our theoretical modelling shows a preference for a homogeneous circumburst medium. Our forward shock parameters are typical for the wider GRB population, and we find that the environment of the burst is characterised by a sub-solar metallicity.
*Conclusions.* Our observations and modelling of GRB 220627A do not suggest that a different progenitor compared to the progenitor of normal long GRBs is required. We find that more observations of ultra-long GRBs are needed to determine if they form a separate population with distinct prompt and afterglow features, and possibly distinct progenitors.

## 1. Introduction

Gamma-ray bursts (GRBs) are flashes of γ-rays lasting from milliseconds to hours typically with isotropic γ-ray luminosities of $\sim 10^{51} - 10^{53}$ erg s$^{-1}$, making them the most luminous explosions observed in the Universe (Zhang 2018). GRBs have traditionally been separated into long and short bursts based on an observed bimodality in their duration distribution, with 2 s taken as the dividing line (Kouveliotou et al. 1993). Short bursts are thought to be caused by the coalescence of two compact objects involving a neutron star, while long bursts are thought to result from the collapse of a massive Wolf-Rayet star, though recent observations have shown there is an overlap between these two classes (e.g. GRB 211211A; Rastinejad et al. 2022; Yang et al. 2022).

A very small number of bursts have been detected with extremely long durations greater than 1000 s (e.g. GRBs 091024A, 101225A, 111209A, 130925A; Gruber et al. 2011; Virgili et al.

⋆ Corresponding author e-mail: `DWTSIM002@myuct.ac.za`

2013; Thöne et al. 2011; Levan et al. 2014; Gendre et al. 2013; Stratta et al. 2013; Evans et al. 2014; Piro et al. 2014). Some authors have suggested that these GRBs form a distinct class of so-called ultra-long GRBs whose progenitors may be different with respect to the collapsar model (Gendre et al. 2013; Levan et al. 2014). Levan et al. (2014) studied three ultra-long GRBs – GRB 101225A, GRB 111209A, and GRB 121027A – and found that they had very similar long-lasting X-ray emission with flares and were situated close to the cores of highly star-forming dwarf galaxies. The extremely long duration of these bursts led them to conclude that their central engines were active for much longer than normal long GRBs, and therefore their progenitors may be blue supergiant stars (Mészáros & Rees 2001; Nakauchi et al. 2013) which have much larger radii than the compact Wolf-Rayet stars that are commonly regarded as the progenitors of long GRBs (Woosley & Bloom 2006). This was the preferred explanation by Gendre et al. (2013) for the extremely long duration of ~25000 s for GRB 111209A, where the additional mass from the outer layers of such a star can power the central engine for much longer, leading to longer-duration γ-ray emission.

Zhang et al. (2014), however, pointed out that not all bursts that had been claimed as ultra-long were actually ultra-long in γ-rays (e.g. GRB 121027A). Instead, the long duration of highly variable X-ray light curves had been used to infer the ultra-long duration (Levan et al. 2014). *Swift* observations of X-ray flares and internal plateaus have shown that the central engine duration is much longer than $T_{90}$ values suggest. Zhang et al. (2014) therefore used X-ray data in addition to γ-ray data to derive central engine durations for 343 GRBs, and found that 21.9% of GRBs have durations $t_{\rm burst} \gtrsim 10^3$ s, and 11.5% have $t_{\rm burst} \gtrsim 10^4$ s. The inference is that ultra-long GRBs may be the tail of a single long GRB population and do not require a separate, blue supergiant progenitor even though the data do not exclude it. Additionally, observations suggest that the afterglow properties of ultra-long GRBs are not different to those of other, classical, long GRBs (Virgili et al. 2013). In order to assess the claim that ultra-long GRBs form a distinct population compared to other long GRBs, further multi-wavelength observational evidence encompassing afterglow and host galaxy properties is needed.

On 2022 June 27, the *Fermi* Gamma-ray Burst Monitor (GBM; Meegan et al. 2009) was triggered by two events separated by 956 s, leading to the speculation that GRB 220627A was a gravitationally lensed or ultra-long GRB with a duration of ~1000 s (Roberts et al. 2022). The Large Area Telescope (LAT; Atwood et al. 2009) aboard *Fermi* detected high-energy photons coincident with the first GBM trigger and localised the first burst to a 0.2 degree-radius error circle (di Lalla et al. 2022), which led to the eventual identification of the optical and X-ray afterglows at ~0.8 days post-trigger. Here we report the results of our observational follow-up campaign and place GRB 220627A in the context of the ultra-long GRB population.

We note that there is no universally agreed-upon definition for ultra-long GRBs, so we defer to adopting the convention that ultra-long bursts satisfy $T_{90} \gtrsim 1000$ s, as used by Lien et al. (2016). We report all uncertainties at the 1σ level unless stated otherwise, and all magnitudes in the AB system. We follow the conventions with $F_\nu \propto \nu^\beta t^\alpha$ and $N_E(E) \propto E^{-\Gamma}$, and adopt a Lambda cold dark matter (ΛCDM) cosmology with $\Omega_m = 0.31$, $\Omega_\Lambda = 0.69$, and $H_0 = 68$ km s$^{-1}$ Mpc$^{-1}$ (Planck Collaboration et al. 2016). The first *Fermi*/GBM trigger occurred at 21:21:00.09 UT. We take this time as $T_0$ for GRB 220627A, and reference all other observations with respect to this time.

## 2. Observations

### 2.1. MeerLICHT optical afterglow discovery

MeerLICHT is a 65-cm aperture fully-robotic optical telescope located at the South African Astronomical Observatory (SAAO) site in Sutherland, South Africa (Bloemen et al. 2016). The primary science goal behind MeerLICHT is to provide simultaneous optical coverage of the radio sky as observed by the 64-antennae MeerKAT radio array, which is also located within the arid Karoo region of the South African interior. MeerLICHT's wide field-of-view of 2.7 deg$^2$ (98.6' × 98.6') and robotic operation make it suitable for the discovery of new transients and follow-up of events with large error boxes on the plane of the sky, such as gravitational wave (GW) events and poorly-localised GRBs, while its six filters (SDSS *ugriz* plus a wide *q* filter spanning 440–720 nm, roughly equivalent to *g + r*) make it suitable for multi-colour monitoring of transients.

MeerLICHT began an observing programme in June 2021 to follow-up GRBs detected by the *Swift* and *Fermi* missions. For *Swift* bursts the aim is to start observing Burst Alert Telescope (BAT) error boxes as soon as possible after a trigger, regardless of whether an X-ray or optical counterpart is discovered by *Swift*. For *Fermi* bursts the aim is to observe a cumulative probability of at least 70% of the GBM HEALPix skymaps with 80 or fewer telescope pointings, an observing criterion which naturally targets brighter bursts since they are better-localised. Observations are automatically triggered for bursts if they can begin within five hours of a trigger, with manual scheduling available for specific cases such as LAT-detected bursts or bursts with a high scientific interest.

GRB 220627A triggered automatic observations with MeerLICHT starting 21 minutes after the trigger at 21:42:23 UT, when the fields became visible given the telescope's observing constraints. The first GBM trigger had a 90% (50%) error area of 216.5 (52.7) deg$^2$, while the second trigger had a 90% (50%) error area of 213.8 (32.5) deg$^2$. A total of 52 fields encompassing a cumulative probability of 71% of the first trigger GBM skymap (see Fig. 1) were observed during the 1.6 hours that MeerLICHT was observing. All observations consisted of 60 s *q*-band exposures, since the *q*-band is our most sensitive band and also has the most complete set of archival reference images. Our automatic scheduling aims to target the highest probability fields first, but due to the western fields setting first only the most eastern fields were scheduled. All 52 fields were observed at least once in the *q* band, while ten of the most eastern fields were observed twice encompassing a cumulative probability of 7.5%. Two of these ten fields were observed three times. The MeerLICHT transient detection pipeline was used to identify two afterglow candidates which were reported via the GCN within six hours of the first trigger (Groot et al. 2022).

The next morning at 07:00:40 UT, a *Fermi*/LAT GCN circular indicated that high-energy γ-rays had been detected along with a 0.2 degree radius localisation - a much smaller error box than the GBM error box (di Lalla et al. 2022). The LAT error box straddled two MeerLICHT fields which had not been observed on the previous night, ruling out our two initial afterglow candidates. We scheduled 2 × 300 s observations in the *q*-band for these fields in order to search for an optical afterglow within the LAT error box. Observations began just after the evening twilight under new moon conditions at 17:12:15 UT, at approximately 0.83 days post-trigger. *Swift*/XRT target of opportunity (ToO) observations identified 15 X-ray sources[1] within the LAT error box that may have been associated with GRB 220627A (Evans & Swift Team 2022), most having error regions with radii smaller than 8″. Among the three most promising[2] candidates (green in Fig. 1), only one source was uncatalogued. Our two MeerLICHT images at 0.84 and 0.87 days showed a new optical transient within the error box of this source at coordinates $\alpha = 13^h25^m28.49^s$, $\delta = -32^d25^m33.31^s$ (ICRS). The brightness of $q = 21.25 \pm 0.09$ mag in our first exposure, and the lack of any source at the same position down to $q > 21.71$ in an archival image from two months earlier made it likely that this source was indeed the optical afterglow to GRB 220627A (de Wet et al. 2022). The source was confirmed as the afterglow through spectroscopic observations with MUSE mounted on the VLT (see Sect. 2.5 below, and Izzo et al. 2022).

---
[1] See the list of sources at the UK Swift Science Data Centre website.

### 2.2. Prompt emission

The duration of the burst associated with the first GBM trigger was $T_{90} = 136.71 \pm 1.28$ s, while that of the burst associated with the second trigger was $T_{90} = 126.98 \pm 8.84$ s, as taken from the online Fermi GBM Burst Catalog (von Kienlin et al. 2020). The 10–1000 keV fluence for the first and second triggers (from the Burst Catalog) are $(4.54 \pm 0.01) \times 10^{-5}$ erg cm$^{-2}$ and $(1.08 \pm 0.02) \times 10^{-5}$ erg cm$^{-2}$, resulting in a total fluence of $(5.62 \pm 0.02) \times 10^{-5}$ erg cm$^{-2}$ across both episodes. At the burst redshift of $z = 3.08$ (Sect. 2.5) this results in a high isotropic $\gamma$-ray energy of $E_{\gamma,\mathrm{iso}} = (4.81 \pm 0.02) \times 10^{54}$ erg. Among the ~130 GRBs with known redshift listed in the 10-year Fermi-GBM Gamma-Ray Burst Spectral Catalog (Poolakkil et al. 2021), only two bursts have a larger isotropic $\gamma$-ray energy: GRBs 090323 and 160625B. Both emission episodes were also detected by *Konus-Wind* (Aptekar et al. 1995). The *Konus-Wind* light curve had a similar structure to the GBM light curve, though it was reported that a weak emission tail may have been present in the soft energy bands (20–100 keV) lasting up until $T_0 + 3700$ s (Frederiks et al. 2022).

GeV photons were detected by *Fermi*/LAT coincident with the first GBM trigger but not with the second one (di Lalla et al. 2022). The detection of GeV photons makes GRB 220627A the first ultra-long (with duration > 1000 s) GRB detected at these energies. The prompt emission from GBM and LAT was studied in detail by Huang et al. (2022). They found that the time-integrated spectrum for the first emission episode is best described by a cutoff power law plus a power law component, where the power law component (with $\Gamma = 1.92 \pm 0.05$) is necessary to account for the detected LAT photons. The second emission episode is best described by a cutoff power law alone. The cutoff energy for the first and second episodes are $E_c = 286.96 \pm 38.00$ keV and $E_c = 248.64 \pm 66.58$ keV, while their low-energy photon indices are $\Gamma = 0.73 \pm 0.10$ and $\Gamma = 1.06 \pm 0.11$, respectively. Huang et al. (2022) argue that the gravitational-lensing scenario to explain the two triggers can be ruled out at the 5.1$\sigma$ level based on the non-detection of any LAT photons coincident with the second trigger. Furthermore, they place a lower limit on the bulk Lorentz factor of $\Gamma \geq 300$ by using the optical depth to pair production method along with the highest energy LAT photon with an energy of 15.3 GeV.

In order to quantify the duration of GRB 220627A across both emission episodes, we obtained the CSPEC data from the NASA HEASARC data archive for the NaI detectors where the burst position was within 50° of the detector normal – these were naturally the detectors with the strongest detected signal. For trigger 1 these were detectors $n0$, $n1$, $n3$ and $n6$, while for trigger 2 these were $n3$, $n4$, $n6$, $n7$, and $n8$. We considered photons with energies between 8 keV and 1 MeV and rebinned the count-rate light curve for each detector in 4.096 s time bins, fitting the background with polynomials of varying order in three time intervals for each detector: before the first trigger, between both triggers, and after the second trigger. These time intervals correspond to $[-250.0, -60.0]$, $[300.0, 900.0]$ and $[1200.0, 1500.0]$ s with respect to $T_0$. The summed count rate light curve for all seven detectors is shown in Fig. 2. We determine a burst duration of $T_{90} \approx 1092$ s as the time difference between the times when the cumulative background-subtracted count rate was between 5% and 95% of its total value. We note that this duration depends on the time binning used and the detectors included in the summed light curve, and hence should be regarded as approximate.

The spectral hardness versus duration diagram has long been used as a means of categorising bursts into long and soft or short and hard classes (Kouveliotou et al. 1993), notwithstanding the fact that the burst duration is energy and detector-dependent (Qin et al. 2013; Bromberg et al. 2013). For both emission episodes we calculated the hardness ratio as the ratio of the integrated flux in the 50–300 keV to 10–50 keV energy ranges, making use of the best-fit parameters from the cutoff power law fits performed by Huang et al. (2022). Figure 3 compares the position of both emission episodes in the hardness-duration diagram with the sample of triggers from the Fourth *Fermi*/GBM Gamma-Ray Burst Catalog (von Kienlin et al. 2020). We see that both emission episodes sit in the long and soft portion of the diagram, providing support for a collapsar origin.

### 2.3. X-rays

The *Swift* X-Ray Telescope (XRT; Burrows et al. 2005) conducted ToO observations of the LAT error box in order to search for X-ray afterglow candidates (Evans & Swift Team 2022). The MeerLICHT identification of an optical source coincident with one of the potential X-ray afterglow candidates (see Sect. 2.1) confirmed this source as the X-ray afterglow, along with evidence of X-ray fading. A total of three orbits of Photon Counting mode (PC) observations of the afterglow were acquired at mid-times of 0.52, 1.59 and 4.62 days post-trigger. We obtained the X-ray count-rate light curve and spectrum from the online *Swift*/XRT GRB Catalogue hosted on the UK Swift Science Data Centre (UKSSDC) website (Evans et al. 2007, 2009). We fitted the photon spectrum with a photoelectrically absorbed power law model (tbabs*ztbabs*pow) in Xspec version 12.12.1, fixing the Galactic hydrogen column density at $N_\mathrm{H}^\mathrm{Gal} = 4.90 \times 10^{20}$ cm$^{-2}$ (Willingale et al. 2013) and fixing the source redshift to $z = 3.08$. From our fit we derive a host column density of $N_\mathrm{H}^\mathrm{host} = 3.75^{+7.83}_{-3.74} \times 10^{22}$ cm$^{-2}$ and a photon index of $\Gamma = 1.73^{+0.78}_{-0.51}$ with a C-statistic of 14.7 for 19 degrees of freedom. The unabsorbed 0.3–10 keV flux was $1.05^{+0.28}_{-0.24} \times 10^{-12}$ erg cm$^{-2}$ s$^{-1}$, from which we derive an unabsorbed counts-to-flux conversion factor of $3.60 \times 10^{-11}$ erg cm$^{-2}$ count$^{-1}$. Using this value along with a spectral index of $\beta_\mathrm{X} \equiv 1 - \Gamma_\mathrm{X} \approx -0.73$ we created a 1 keV X-ray light curve. The light curve consisted of two detections at 0.52 and 1.59 days, and a 3$\sigma$ upper limit at 4.62 days (see Fig. 4).

### 2.4. Optical/near-infrared photometry

Following confirmation of the afterglow detection, we obtained 600 s exposures in each of the $q$, $g$, $r$, and $i$ bands with Meer-

---
[2] After source detection, *Swift* sources are given a quality flag (Good, Reasonable or Poor) which indicates how likely a source is to be real.

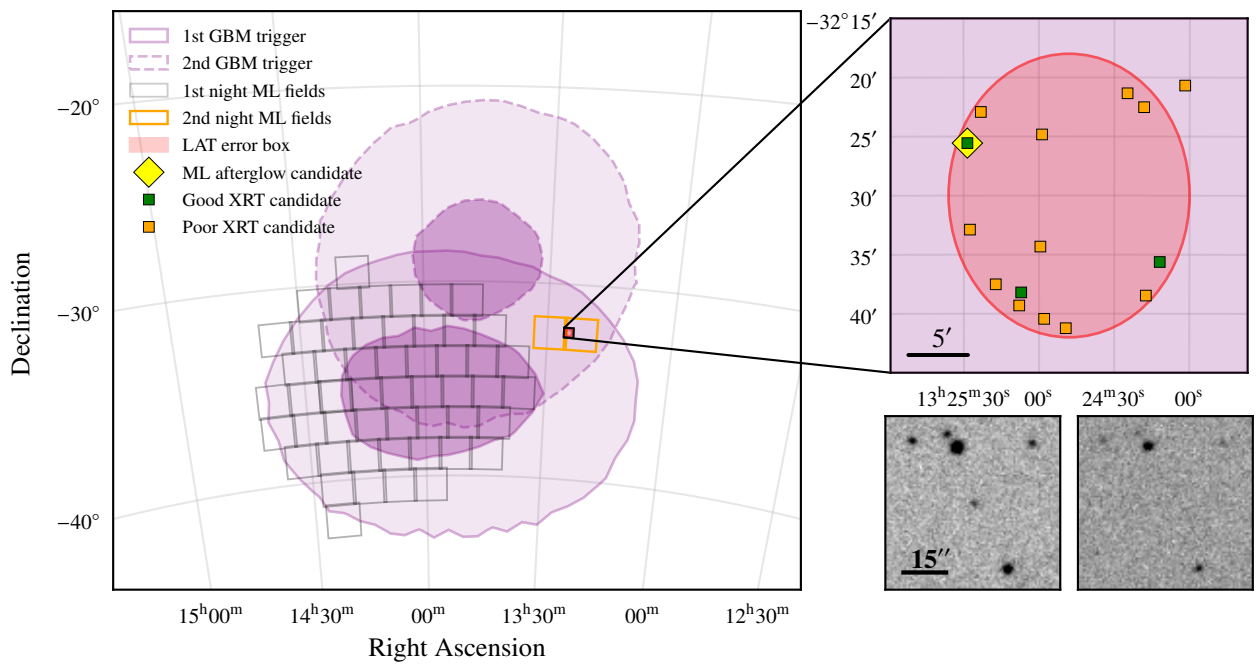

**Fig. 1.** Localisation of GRB 220627A. *Left:* 50% and 90% uncertainty regions for both GBM triggers associated with GRB 220627A are outlined in purple. A total of 52 MeerLICHT fields (grey boxes) were automatically scheduled for observations immediately following the first trigger. Following the announcement of the *Fermi*/LAT localisation region (red circle), two MeerLICHT fields (orange boxes) were scheduled for observations the following night. *Top right: Swift*/XRT ToO observations identified 15 X-ray sources within the LAT error box. MeerLICHT identified a new optical transient within the error box of the third 'good' XRT source (yellow diamond). *Bottom right:* First MeerLICHT detection of the new transient (left), and an archival image taken two months previously on 2022 April 26 (right). Both thumbnails are $1' \times 1'$ in dimension.

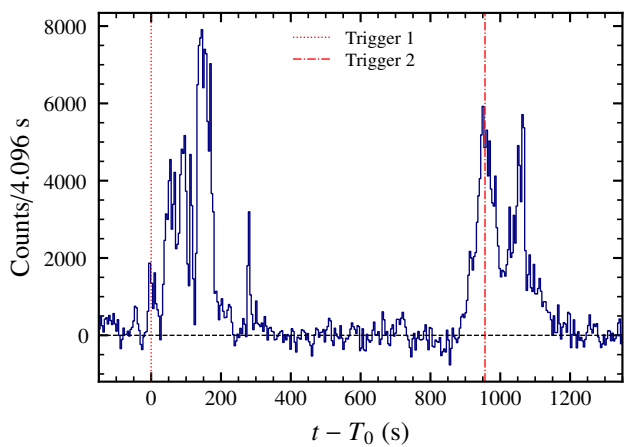

**Fig. 2.** Prompt emission light curve created through summing the background-subtracted light curves from seven GBM NaI detectors. We measure $T_{90} \approx 1092$ s. Vertical lines mark the *Fermi*/GBM trigger times.

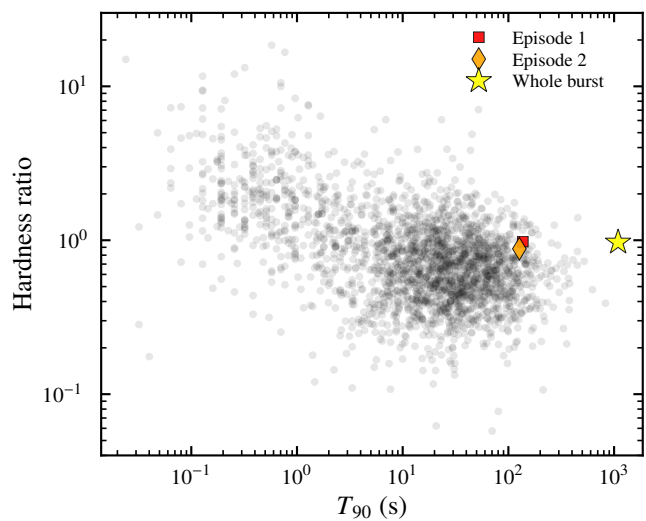

**Fig. 3.** Hardness-duration diagram for both GRB 220627A emission episodes in comparison with the sample from the Fourth *Fermi*/GBM Gamma-Ray Burst Catalog (von Kienlin et al. 2020).

LICHT at ~1.84 days post-trigger. We used the MeerLICHT pipeline (Vreeswijk et al., in prep) to perform standard CCD reduction tasks including calibration, astrometry and point-spread function (PSF) photometry. Due to the faintness of the afterglow in some of our images, we employed a forced photometry routine (developed as part of the MeerLICHT pipeline) to accurately measure the flux and significance of the detections. The afterglow was detected with a high significance ($> 15\sigma$) in the first two $q$-band images at 0.84 and 0.87 days. The afterglow was detected at more than $5\sigma$ significance in the $q$ and $r$ bands at 1.84 days, and at $4.7\sigma$ significance in the $g$ band. The $i$-band observation yielded a low-significance measurement of $2.7\sigma$ which we do not include as a detection.

We obtained $6 \times 600$ s exposures at approximately 0.93 days post-trigger in the $g$, $r$, and $i$ bands with the 1-m SAAO Lesedi optical telescope (Worters et al. 2016) located in Sutherland and equipped with the Mookodi spectrograph and imager (Erasmus et al., in prep). We employed an adapted version of the Meer-LICHT pipeline to perform astrometry and photometry on each image since both telescopes make use of the SDSS *ugriz* filter set. The pipeline produced a catalogue file containing all $5\sigma$ source detections. We detected the afterglow in each image.

We acquired three epochs of imaging in the Bessel *R* filter at 1.13, 5.09, and 360.2 days post-trigger with the European Southern Observatory Very Large Telescope (ESO VLT) UT1 (Antu) equipped with FORS2. We obtained the last epoch approximately one year after the GRB in order to constrain any host galaxy emission long after the afterglow had faded. Each epoch consisted of 6×200 s, 3×200 s and 3×300 s exposures, respectively. We performed photometric calibration in the *R* band using *ri* DELVE DR2 photometry (Drlica-Wagner et al. 2022) of a large number of stars in the field along with the Lupton (2005)[3] transformation equation $R = r − 0.2936 ∗ (r − i) − 0.1439$. We employed the Aperture Photometry Tool (APT; Laher et al. 2012) with a three-pixel radius aperture in an automated mode to extract instrumental magnitudes of all $3\sigma$ point sources within the FORS2 images, and derived image zero-points of $33.18 \pm 0.02$, $33.37 \pm 0.01$, and $33.42 \pm 0.03$ mag in the images from 1.13, 5.09, and 360.2 days, respectively. We detected the afterglow with a brightness of $R = 21.25 \pm 0.02$ and $R = 24.61 \pm 0.10$ mag during the first two epochs, and derived a $3\sigma$ upper limit of $R > 25.92$ mag during the third epoch.

We further obtained two epochs of follow-up observations of GRB 220627A in the $g'r'i'z'JHK$ bands with the Gamma-ray Burst Optical Near-Infrared Detector (GROND; Greiner et al. 2008) mounted at the 2.2-m MPG telescope at the ESO La Silla observatory in Chile. The afterglow was detected in all the optical bands ($g'r'i'z'$) but not in the near-infrared bands ($JHK$) during the first epoch of observations at 2.18 days post-trigger (Nicuesa Guelbenzu et al. 2022). The afterglow was detected in only the $g'$ and $r'$ bands during the second epoch at 3.21 days. The GROND data were reduced using standard PSF photometry through DAOPHOT (Stetson 1987) and IRAF (Tody 1993). The optical data were calibrated to the Pan-STARRS catalogue (Chambers et al. 2016) while the NIR data were calibrated to the 2MASS catalogue (Skrutskie et al. 2006). We present all optical detections of GRB 220627A in Fig. 4.

### 2.5. Optical spectroscopy

Following the detection of the optical afterglow by MeerLICHT, we observed the GRB afterglow region with the Multi-Unit Spectroscopic Explorer (MUSE; Bacon et al. 2010) mounted on UT4 at the ESO VLT. A set of four exposures of 600 s were obtained at 1.17 days post-trigger covering the wavelength range 4800–9300 Å with the corresponding spectral resolution ranging from $R \sim 1800$–3500. Data from each exposure was reduced and stacked using standard *esorex* recipes. Finally, we subtracted the sky background in the reduced data cube using *zap* (Soto et al. 2016). We extracted a spectrum from the MUSE data cube using a circular aperture of 0.6″. The spectrum is characterised by the presence of a broad damped Ly$\alpha$ absorption feature and the presence of several narrow metal absorption lines, including O I, Si II, Si II*, C II, C II*, Si IV, C IV, Al II, Fe II, at the common redshift of $z = 3.084$ (Izzo et al. 2022). An intervening absorber at $z = 2.665$ was also identified by the presence of S IV, Si II, C IV, Fe II, Al II lines (Izzo et al. 2022). The white light image, obtained by integrating over the entire wavelength region covered by MUSE, is shown in Fig. 5 along with the resulting spectrum.

---
[3] See http://classic.sdss.org/dr4/algorithms/sdssUBVRITransform.html.

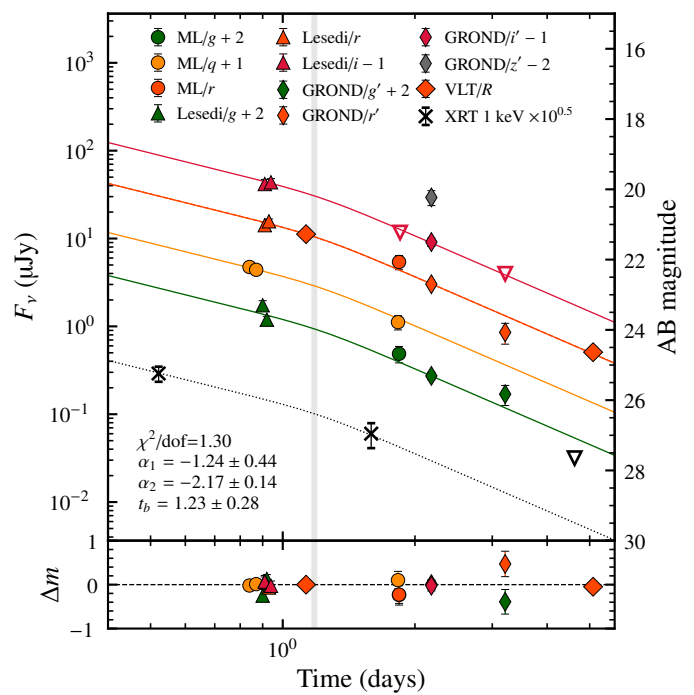

**Fig. 4.** X-ray and optical light curves associated with GRB 220627A, organised by observing band and instrument. MeerLICHT is abbreviated to ML in the legend. The vertical grey region denotes the time of the MUSE spectroscopic observations. We show the broken power law fits to each of the $g$, $q$, $r$ and $i$ bands, where the break time and temporal indices were constrained to be the same across each fit. Residuals are shown in magnitudes in the lower panel. We show the two X-ray detections as crosses and the single upper limit as an upside-down triangle. The dotted line is the optical light curve fit shifted vertically to the X-ray band.

### 2.6. Radio

Radio observations with the ATCA and MeerKAT arrays were obtained at 8 and 44 days post-trigger, respectively (Leung et al. 2022; Giarratana et al. 2022). The radio afterglow of GRB 220627A was detected by ATCA with a flux of ∼0.4 mJy at a frequency of 17 GHz at 7.3 days, while no source was detected by MeerKAT in the L-band at 8.9 and 35.7 days where the image RMS noise was 9 and 14 µJy for each epoch, respectively. We assume a conservative 33% error on the ATCA flux measurement, and the upper limits as three times the RMS noise ($3\sigma$).

All flux measurements and upper limits used in this work are presented in Table 1, where we have converted AB magnitudes to flux densities in µJy. The frequency of each optical filter corresponds to the effective wavelength of that filter, which we obtained from the SVO Filter Profile Service for the VLT and GROND filters, while for MeerLICHT and Lesedi we obtained them internally.

## 3. Results

### 3.1. External forward shock framework

We interpret our afterglow observations in the framework of the synchrotron external forward shock model (Mészáros & Rees 1997; Sari et al. 1998; Chevalier & Li 2000). In this model, the GRB central engine – usually assumed to be a black hole or neutron star – powers an extremely relativistic, collimated outflow that sweeps up mass in the surrounding circumburst medium, forming a shock front behind which electrons are accelerated to a power law distribution in energies with $N(\gamma_e) \propto \gamma_e^{-p}$ for $\gamma_e > \gamma_m$, where $\gamma_m$ is the minimum Lorentz factor of the electrons in the distribution and $p$ is the electron spectral index. The electrons produce synchrotron radiation whose emission spectrum is a multi-segment broken power law characterised by three break frequencies: the frequency associated with the peak of the spectrum, $\nu_m$, corresponding to electrons with the Lorentz factor $\gamma_m$ in the instantaneous electron distribution; the cooling frequency $\nu_c$ corresponding to the Lorentz factor beyond which electrons are cooling efficiently by synchrotron radiation over the lifetime of the system; and the self-absorption frequency $\nu_{sa}$ corresponding to the frequency below which the synchrotron emission is self-absorbed. The ordering and evolution of the spectral breaks with time is dictated by the dynamics of the blast wave. After reverse shock crossing, the decelerating blast wave enters a self-similar regime described by the Blandford-McKee (BM; Blandford & McKee 1976) solution for a spherical, relativistic blast wave expanding into a medium with a density profile $n(r) \propto r^{-k}$. From this solution, the spectral flux density is given as a power law of both time and frequency, from which so-called closure relations can be derived relating the temporal and spectral indices within a given spectral regime (Zhang & Mészáros 2004; Zhang et al. 2006; Gao et al. 2013). We consider both a constant density ISM-like medium ($k = 0$) or a stellar wind medium ($k = 2$) in the relativistic (BM) regime, as in Granot & Sari (2002).

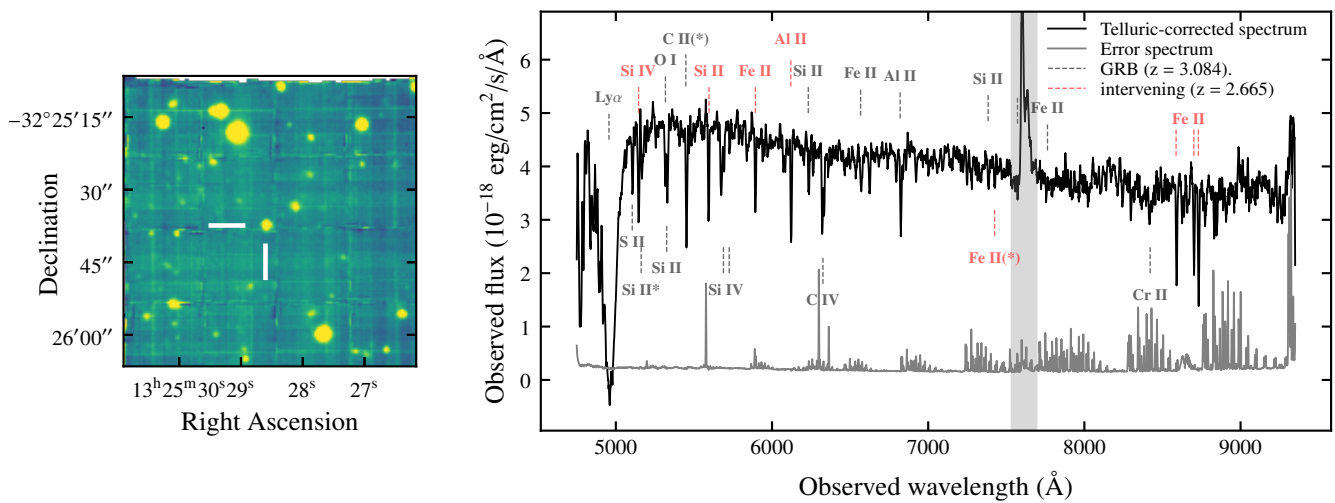

**Fig. 5.** MUSE spectroscopy of GRB 220627A. *Left:* White light image resulting from the final MUSE data cube. The GRB afterglow is clearly visible at the centre of the image. *Right:* Telluric-corrected spectrum (black) and error spectrum (grey) of GRB 220627A extracted from the MUSE data cube using an aperture with a radius of 0.6″. We show identified absorption lines at the GRB redshift ($z = 3.084$) and the intervening system ($z = 2.665$) in grey and red, respectively. The vertical shaded region is heavily affected by telluric lines.

### 3.2. Evidence for a jet break

Optical detections of GRB 220627A are presented in Fig. 4, separated by observing band. To account for the slight differences in observing filters, we shifted the $r$- and $R$-band data to a common frequency of $4.820 \times 10^{14}$ Hz (the GROND $r'$-band effective frequency) using a spectral index of $\beta \approx -1$. This is the approximate spectral index derived from the GROND $r'$, $i'$ and $z'$ bands at 2.18 days, uncorrected for Galactic extinction. We excluded the $g'$ band detection from the spectral fit as there is a clear dip in the spectral energy distribution (SED) in this band compared to the redder bands, due to rest-frame Ly$\alpha$ absorption (see Fig. 5).

We fitted the light curves in each of the $g$, $q$, $r$, and $i$ bands with two analytic functions: a simple power law (PL), and a smoothly broken power law[4] (BPL), constraining the temporal indices and break times (for the BPL) to be the same across each fit. We employed a break smoothness parameter of $\omega = 9$ since smoother breaks result in fits with higher reduced $\chi^2$ values. The BPL fit had a reduced $\chi^2$ value substantially closer to one compared to the simple PL fit (1.30 versus 1.87). Furthermore, the Bayesian information criterion (BIC) for the two fits gives preference to the BPL model (16.09 versus 19.86). Figure 4 shows the best-fit BPL model in each optical band. The pre- and post-break temporal indices are $\alpha_1 = -1.24 \pm 0.44$ and $\alpha_2 = -2.17 \pm 0.14$, with a break time of $t_b = 1.23 \pm 0.28$ days. Our late-time constraint on any host galaxy emission of $R > 25.92$ mag allows us to place an upper limit to any possible host galaxy contribution to the final $R$-band detection at 5.09 days of <30%, providing additional support to the steepening observed in the light curve. The BPL model is also compatible with the X-rays, as the data can accommodate the optical light curve fit shifted to the X-ray band (Fig. 4).

At X-ray and optical frequencies, an achromatic steepening of the afterglow light curves from a 'normal' temporal index of $\alpha \approx -1$ to a post-break index of $\alpha \approx -2$ has often been attributed to the jet break effect. Observationally, jet breaks have been studied extensively (see the sample in Wang et al. 2018). As the blast wave decelerates, a deficit in flux will be observed once $1/\Gamma > \theta_j$, resulting in a steepening of the light curves. If the jet break is purely due to sharp edges of the jet (e.g. a top-hat jet) coming into view rather than due to lateral spreading or a combination of both, the light curves at all frequencies will steepen by $t^{-3/4}$ or $t^{-1/2}$ in an ISM or stellar wind environment, respectively. Rhoads (1999) and Sari et al. (1999) considered sideways expansion of a conical jet and found that the bulk Lorentz fac-in which $F_0$ is the normalising flux level, $\alpha_1$ and $\alpha_2$ are the pre- and post-break temporal indices, $t_b$ is the break time, and $\omega$ is a smoothness parameter.

---

[4] We employ the functional form $F(t) = F_0 \left[ \left( \frac{t}{t_b} \right)^{-\alpha_1 \omega} + \left( \frac{t}{t_b} \right)^{-\alpha_2 \omega} \right]^{-1/\omega}$

**Table 1.** X-ray, optical and radio observations of GRB 220627A used in this work.

| $\Delta t$ (days) | Telescope | Band | Wavelength (nm) | Frequency (Hz) | Flux (µJy) | Uncertainty (µJy) |
|---|---|---|---|---|---|---|
| 0.52 | *Swift*/XRT | 1 keV | - | 2.42e+17 | 0.092 | 0.018 |
| 1.59 | *Swift*/XRT | 1 keV | - | 2.42e+17 | 0.019 | 0.006 |
| 4.62 | *Swift*/XRT | 1 keV | - | 2.42e+17 | < 0.010 | - |
| 0.84 | MeerLICHT | q | 580 | 5.169e+14 | 11.91 | 0.66 |
| 0.87 | MeerLICHT | q | 580 | 5.169e+14 | 11.07 | 0.71 |
| 1.83 | MeerLICHT | q | 580 | 5.169e+14 | 2.81 | 0.52 |
| 1.84 | MeerLICHT | g | 480 | 6.246e+14 | 3.08 | 0.65 |
| 1.84 | MeerLICHT | r | 626 | 4.789e+14 | 5.45 | 1.00 |
| 1.85 | MeerLICHT | i | 765 | 3.919e+14 | < 4.66 | - |
| 0.90 | Lesedi | g | 466 | 6.439e+14 | 10.97 | 1.41 |
| 0.91 | Lesedi | r | 611 | 4.907e+14 | 13.93 | 1.28 |
| 0.91 | Lesedi | i | 758 | 3.954e+14 | 16.60 | 2.14 |
| 0.92 | Lesedi | g | 466 | 6.439e+14 | 7.52 | 0.69 |
| 0.93 | Lesedi | r | 611 | 4.907e+14 | 15.42 | 0.99 |
| 0.94 | Lesedi | i | 758 | 3.954e+14 | 17.38 | 1.76 |
| 1.13 | VLT | R | 642 | 4.666e+14 | 11.59 | 0.21 |
| 5.09 | VLT | R | 642 | 4.666e+14 | 0.53 | 0.05 |
| 360.2 | VLT | R | 642 | 4.666e+14 | < 0.156 | - |
| 2.18 | GROND | g' | 459 | 6.536e+14 | 1.72 | 0.17 |
| 2.18 | GROND | r' | 622 | 4.820e+14 | 3.02 | 0.17 |
| 2.18 | GROND | i' | 764 | 3.924e+14 | 3.63 | 0.40 |
| 2.18 | GROND | z' | 899 | 3.335e+14 | 4.66 | 0.90 |
| 2.18 | GROND | J | 1240 | 2.418e+14 | < 10.00 | - |
| 2.18 | GROND | H | 1647 | 1.820e+14 | < 20.89 | - |
| 2.18 | GROND | K | 2170 | 1.381e+14 | < 47.87 | - |
| 3.21 | GROND | g' | 459 | 6.536e+14 | 1.07 | 0.28 |
| 3.21 | GROND | r' | 622 | 4.820e+14 | 0.85 | 0.23 |
| 3.21 | GROND | i' | 764 | 3.924e+14 | < 1.58 | - |
| 3.21 | GROND | J | 1240 | 2.418e+14 | < 12.02 | - |
| 3.21 | GROND | H | 1647 | 1.820e+14 | < 22.91 | - |
| 3.21 | GROND | K | 2170 | 1.381e+14 | < 47.86 | - |
| 7.3 | ATCA | Ku | - | 1.7e+10 | 400 | 132 |
| 8.9 | MeerKAT | L | - | 1.4e+09 | < 27 | - |
| 35.7 | MeerKAT | L | - | 1.4e+09 | < 42 | - |

**Notes.** The optical data have not been corrected for Galactic or host galaxy extinction. Radio measurements were obtained from GCN circulars (Leung et al. 2022; Giarratana et al. 2022). We provide the central wavelength of the optical/NIR observing bands, and list all upper limits at the $3\sigma$ level.

tor of the jet decreases exponentially after $1/\Gamma > \theta_j$, producing a steeper post-jet break decay. For the spectral break ordering with $\nu_a < \nu_m < \nu_c$ (usually relevant when a jet break occurs), Sari et al. (1999) found that the light curves decay as $t^{-p}$ for $\nu > \nu_m$, $t^{-1/3}$ for $\nu_a < \nu < \nu_m$, and $t^0$ for $\nu < \nu_a$. Numerical simulations, however, have shown that sideways expansion should not contribute before $\Gamma$ has decreased considerably, though post-jet break decay is predicted to be steeper than with the edge effect only so that $t^{-p}$ may be a reasonable approximation (van Eerten & MacFadyen 2012b; Granot & Piran 2012; Zhang & MacFadyen 2009).

The passage of the cooling frequency through the optical bands can in principle lead to a temporal break in optical light curves. For both an ISM and wind medium undergoing slow cooling the difference in spectral index between the spectral segments either side of the cooling break is $\Delta\beta = 0.5$. Interpolating and extrapolating the $R$-band light curve fit to the same time as the two X-ray detections, we find that the optical to X-ray spectral index either side of the break in the light curve is $\beta_{O,X} \approx -0.9$, indicative of negligible spectral evolution between the optical and X-ray bands during this time. This spectral slope is also consistent with the optical in-band spectral index of $\beta_O = -0.91$ at 2.18 days, indicating that both bands may lie on the same spectral segment. The difference in decay rate between the regimes either side of the cooling break is $|\Delta\alpha| = 0.25$, so that the passage of $\nu_c$ should not lead to a steepening of more than $\Delta\alpha = -0.25$. We also expect the temporal break to be chromatic, that is it occurs at different times in different observing bands. Our light curves steepen by $\Delta\alpha = -0.93 \pm 0.46$, which is too large for the passage of the cooling break but is consistent with a jet break due to the edge effect ($\Delta\alpha = -0.75$ in an ISM medium) or lateral spreading ($t^{-p}$). Furthermore, the post-break decay index of $\alpha \approx -2.2$ is difficult to reconcile within the standard closure relations. The steepest decay rate is expected in the regime with $\nu_m < \nu < \nu_c$ within a stellar wind scenario, where $\alpha = (1 - 3p)/4$. An extreme electron power law index of $p = 3$ would result in $\alpha = -2$, which is still shallower than our measured value of $\alpha \approx -2.2$. It is therefore likely that the break in our light curve is a jet break. We investigate the jet-break scenario further via theoretical modelling in Sect. 4.

*3.3. Broadband temporal and spectral considerations*

The GROND optical spectral energy distribution (SED) at 2.18 days has a spectral slope of $\beta_O = -0.91 \pm 0.16$ as derived from the $r'$, $i'$ and $z'$ bands (The *JHK* near-infrared bands did not constrain the spectral slope) corrected for Galactic extinction using the Milky Way extinction curve from Fitzpatrick (1999) with $R_V = 3.1$ and $A_V = 0.13$ mag for the GRB line of sight (Schlafly & Finkbeiner 2011). We note that these observing bands correspond to rest-frame ultraviolet (UV) wavelengths at the redshift of GRB 220627A ($z = 3.08$), so that the intrinsic spectral slope may be shallower than this value if there is significant host-galaxy extinction. The X-ray spectral index from our fit in Sect. 2 was $\beta_X = -0.73^{+0.57}_{-0.77}$. We calculate an optical to X-ray spectral index at the time of the second X-ray detection at 1.59 days by interpolating our $r$-band fit to this time. We take the mean relative error in our $r$-band flux measurements as the error on the interpolated $r$-band flux, approximately 10%. We measure a spectral index of $\beta_{O,X} = -0.93 \pm 0.08$, which is consistent with the GROND spectral index of $\beta_O = -0.91 \pm 0.16$. With even moderate host-galaxy extinction, we would expect this to set an upper limit on the optical to X-ray spectral index, such that a value $\beta \approx -1$ is likely from optical to X-ray frequencies.

Due to a paucity of data, we are not able to estimate the location or evolution of the cooling break, $\nu_c$. It is therefore difficult to perform an accurate closure relation analysis to determine the circumburst medium density profile and electron energy spectral index $p$. Nevertheless, if we start with the assumption that $p \approx 2.2$ based on a post-jet break decay of $\alpha_2 = -2.17 \pm 0.14$, we can draw some broad conclusions. In the slow cooling regime ($\nu_m < \nu_c$) we have spectral slopes of $\beta = (1 - p)/2$ and $\beta = -p/2$ below and above $\nu_c$, respectively. Assuming $p \approx 2.2$, we would have $\beta \approx -0.6$ and $\beta \approx -1.1$ below and above $\nu_c$. Our optical to X-ray spectral index of $\beta \approx -1$ therefore appears more consistent with being in the spectral regime above $\nu_c$. With $p = 2.2$, the expected light curve decay rate in this spectral regime is $\alpha = (2 - 3p)/4 = -1.15$, which is fully consistent with the pre-break temporal index of $\alpha = -1.24 \pm 0.44$ from our optical light curve fit. In an ISM environment, $\nu_c$ moves to lower frequencies as $t^{-1/2}$, while in a stellar wind environment $\nu_c$ is expected to rise as $t^{1/2}$ (Granot & Sari 2002). The most plausible[5] scenario is that we are in an ISM environment where $\nu_c$ has already moved below the optical observing bands at the time of our measurement of $\beta_{O,X}$ at 1.59 days.

The radio data consists of a single detection at 7.3 days at 17 GHz, and two non-detections in the L-band (1.4 GHz) at 8.9 and 35.7 days. If we assume that the MeerKAT upper limit at 8.9 days also holds at 7.3 days, we can constrain the spectral slope between the L-band and 17 GHz (Ku band) to $\beta_{L-Ku} \gtrsim 1.07$ at 7.3 days. This implies that the spectrum is likely synchrotron self-absorbed at GHz frequencies.

Since we cannot precisely constrain the location of $\nu_m$, $\nu_c$, or $\nu_a$, there are likely to be extensive model degeneracies when performing theoretical modelling, as shown in Sect. 4.

*3.4. MUSE spectroscopy*

The huge photon flux emitted by early GRB afterglows allows us to investigate the composition of the immediate environment surrounding the GRB progenitor star (Savaglio et al. 2003; Jakobsson et al. 2004; Fynbo et al. 2006; Prochaska et al. 2007; Heintz et al. 2018; Bolmer et al. 2019; Saccardi et al. 2023). This is one of the main motivations to get an optical/near-IR GRB spectrum as soon as an optical counterpart has been identified. Early GRB afterglow spectra have shown the presence of neutral and singly-ionised low-excitation metal absorption lines, such as O I, C II, Si II, S II, Fe II, Ni II, Zn II, Al II, and Al III (Vreeswijk et al. 2007). In several cases, fine structure and meta-stable levels of Si II and Fe II have also been detected (D'Elia et al. 2007). The study of these absorption features can provide us with information on the metallicity of the immediate GRB region, while from the estimate of the column densities $N_X$, with $X$ being the element under consideration, one can also obtain a rough estimate of the extinction $A_V$ along our line of sight (De Cia et al. 2016). Due to their large redshifts, GRB afterglows are therefore one of the most important tools for investigating the evolution of the properties of the ISM in long GRB host galaxies, and consequently, in star-forming galaxies over a large redshift range.

The HI column density in GRB 220627A is derived by modelling the broad Ly$\alpha$ absorption trough with a Voigt profile using VoigtFit (Krogager 2018). This code convolves the intrinsic model with the spectral resolution and determines the redshift $z$, broadening parameter $b$, and column density $N$, of the input lines. Due to the substantial column density of HI, the Lorentzian wings dominate the line profile allowing for accurate estimates of the column density. We derive a best-fit $\log(N_{HI}/cm^{-2}) = 21.15 \pm 0.05$ (see Fig. 6).

We also attempt to measure the column densities of the main elements identified in the MUSE spectrum (Fig. 7). To this aim, and given the spectral resolution of MUSE, we have used the curve of growth (CoG) methodology, which provides a relation between the equivalent width (EW) of ISM lines and the corresponding column densities (Spitzer 1998). For low EW values (<0.1 Å), the EW is directly proportional to the column density, while for larger values the situation is complicated by saturation effects. The dependence of the EW from the column density is also related to the Doppler parameter $b$, which could not be very well constrained with the spectral resolution provided by MUSE. However, if we identify at least two absorption lines originating from the same electronic transition of the same ion (such as Si II 1260, 1808 Å), we can roughly estimate the $b$ value from their EWs, given that these lines have the same column density. Then, under the assumption that the curve of growth of ions with similar excitation potential is described by the same Doppler parameter, we can estimate their column density from their rest-frame EW. Figure 8 shows the results of our methodology. The majority of the lines used for the analysis are consistent with the CoG determined using Si II lines, with the interesting exception of S II. We also note that we have not applied any correction for dust depletion, which can considerably affect ions such as Fe II.

Following prescriptions in Savaglio et al. (2012), we estimate metal abundances assuming that ions with ionisation potentials just above the ionisation energy of hydrogen constitute the dominant ionisation level, and finally give an estimate for the element abundance. We also do not include any ionisation correction given that this is negligible in Damped Lyman $\alpha$ (DLA) systems (Jenkins 2009), and especially when $\log(N_{HI}/cm^{-2}) > 10^{20.5}$ (Péroux et al. 2007), as with GRB 220627A. To determine the metallicity, we derive the relative metal to hydrogen abundances in the GRB afterglow spectrum assuming Solar values provided by Asplund et al. (2009). The metallicity is then determined as $[X/H] = \log(N_X/N_H)_{GRB} - \log(N_X/N_H)_\odot$ for each element identified in the GRB spectrum. In Table 2 we report our final estimates of [X/H] for each ion identified in the GRB spectrum. In

---

[5] We cannot conclusively rule out a wind environment with $\nu_c$ below the optical since the optical light curve spans less than a single decade in time which corresponds to a change in $\nu_c$ by a factor of ~3 given a $t^{1/2}$ evolution.

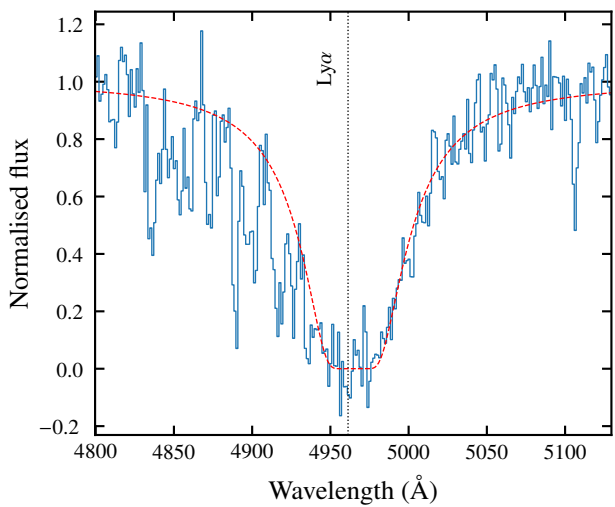

**Fig. 6.** Lyα absorption feature in the MUSE spectrum. The best-fit Voigt profile is shown in red, with $\log(N_{HI}/\text{cm}^{-2}) = 21.15 \pm 0.05$.

**Table 2.** Metallicity computed from the MUSE spectrum and using the CoG method for metal lines.

| Ion | $\lambda_{\text{obs}}$ (Å) | $EW_\lambda$ (Å) | $\log N_X$ (cm$^{-2}$) | [X/H] |
|---|---|---|---|---|
| S II 1253 | 5119.7 | 0.62±0.11 | 15.08±0.09 | -1.19±0.06 |
| Si II 1260 | 5146.4 | 2.55±0.24 | 14.99±0.10 | -1.67±0.08 |
| Si II* 1264 | 5164.1 | 1.20±0.09 | 13.61±0.10 | - |
| O I 1302 | 5316.4 | 1.69±0.08 | 15.35±0.11 | -2.49±0.13 |
| Si II 1304 | 5326.3 | 1.60±0.06 | 14.99±0.10 | -1.67±0.08 |
| C II 1334 | 5448.6 | 1.73±0.10 | 14.95±0.10 | -2.63±0.13 |
| C II* 1335 | 5454.4 | 2.00±0.13 | 15.33±0.10 | - |
| Fe II 1608 | 6568.0 | 0.93±0.08 | 14.38±0.09 | -2.37±0.12 |
| Al II 1670 | 6822.1 | 1.85±0.09 | 13.45±0.09 | -2.15±0.11 |
| Si II 1808 | 7384.4 | 0.73±0.15 | 14.99±0.10 | -1.67±0.08 |

Fig. 9 we compare our results with the distribution of absorption-derived GRB and QSO-DLA metallicities (Rafelski et al. 2012; Thöne et al. 2013; Bolmer et al. 2019; Saccardi et al. 2023), where [X/H] is mainly derived from the S II and Si II ions, which are also the faintest detected lines in the GRB 220627A spectrum, see also Fig. 7. We further notice that for the estimate of the metallicity from S II lines, we have used the S II 1253 Å transition, given that the other line visible in the spectrum, S II 1250 Å, is strongly blended with Si IV 1394 Å from the intervening absorber. The environment of GRB 220627A is characterised by a typical sub-solar GRB afterglow metallicity.

Prochaska (2006) discussed this methodology comprehensively and concluded that column densities obtained using low-resolution spectra are likely to be underestimated due to instrumental line broadening and consequent blending of more intrinsically narrow absorption components. This is particularly true when metal absorption lines display heavily saturated profiles. In the case of GRB 220627A, we are also limited by the low spectral resolution provided by MUSE, which is not suitable to study ISM lines with smaller Doppler parameters. With these prescriptions, the metallicity values determined above must be considered as lower limits, especially for oxygen and carbon that can be heavily saturated. However, we also note that in the very few cases where GRB afterglows have been studied with both high- and low- resolution data, the results obtained using analysis based with Voigt-line profile fits and with the CoG method were fully in agreement (D'Elia et al. 2011).

## 4. Theoretical afterglow modelling

We perform theoretical afterglow modelling to determine the physical blast wave parameters of the forward shock, with the aim of establishing whether the afterglow properties of this ultra-long GRB are unusual compared to the wider GRB population.

### 4.1. Model

We make use of the `ScaleFit` software package to perform theoretical modelling of our broadband afterglow dataset (Ryan et al. 2015; Aksulu et al. 2020, 2022). `ScaleFit` is based on the `BoxFit` set of high-resolution hydrodynamic simulations of GRB explosions (van Eerten et al. 2012). `Boxfit` interpolates between a series of compressed data sets of hydrodynamical jet simulations and employs scale invariance between the blast wave energies and circumburst medium densities to cover the space of afterglow parameters, rapidly computing light curves and spectra using a linear radiative transfer module. `Scalefit` builds on this by making use of simple scaling laws between a set of scale-invariant characteristic quantities to derive the full set of spectral parameters that determine the observed emission (van Eerten & MacFadyen 2012a). The characteristic quantities are computed and tabulated directly from the `BoxFit` simulations, allowing for nearly instantaneous calculation of afterglow light curves and spectra given an arbitrary set of blast wave parameters. The initial conditions for the `BoxFit` simulations are the Blandford-McKee radial profile truncated to the opening angle of the jet $\theta_j$. `ScaleFit` is therefore only suitable for the deceleration phase after reverse shock crossing and so cannot be used to model energy injection, plateaus or flares. The jet break and non-relativistic (Sedov-Taylor) transitions are both well-handled by `ScaleFit` (van Eerten & MacFadyen 2012a). `ScaleFit` can be used to model both an ISM or stellar wind medium, and accepts ten free parameters: explosion-related parameters including the isotropic-equivalent blast wave energy $E_{K,\text{iso}}$, circumburst medium density $n_0$, and opening angle of the jet $\theta_j$; shock microphysics parameters including the electron energy distribution spectral index $p$, the fraction of accelerated electrons within the shock $\xi_N$, and the fractions of shock internal energy partitioned to electrons and magnetic fields, $\epsilon_e$ and $\epsilon_B$, respectively; and observer-related parameters including the burst redshift $z$, luminosity distance $d_L$, and observer angle with respect to the jet axis $\theta_{\text{obs}}$.

We use the `emcee` Python package (Foreman-Mackey et al. 2013) to implement a Markov chain Monte Carlo (MCMC) exploration of our `ScaleFit` model parameter space. An MCMC analysis allows for correlations (and degeneracies) between parameters to be readily identified. Furthermore, the uncertainty on each parameter can be determined through marginalisation of the posterior distribution. `emcee` offers advantages over traditional MCMC samplers because its multiple 'walkers' moving through the parameter space make it affine-invariant, with the result that it is unaffected by covariances between parameters. For GRB afterglow modelling, if a large majority of the measured data points are in a single spectral regime (e.g. at optical and X-ray frequencies) the location of the spectral breaks will not be well constrained, and so there are likely to be strong correlations between the $E_{K,\text{iso}}$, $n_0$, $\epsilon_e$ and $\epsilon_B$ parameters.

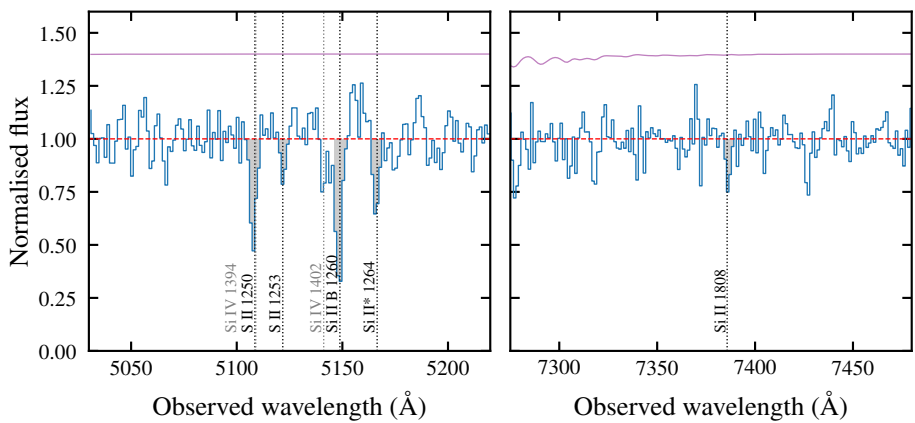

**Fig. 7.** Absorption lines of S II and Si II used in the CoG analysis. We show in grey the Si IV lines from the intervening absorber at $z = 2.665$. The Si IV 1394 line from the absorber is blended with the Si II 1250 line. The purple horizontal line denotes the telluric correction applied to the spectrum, shifted vertically up by 1.4 flux units.

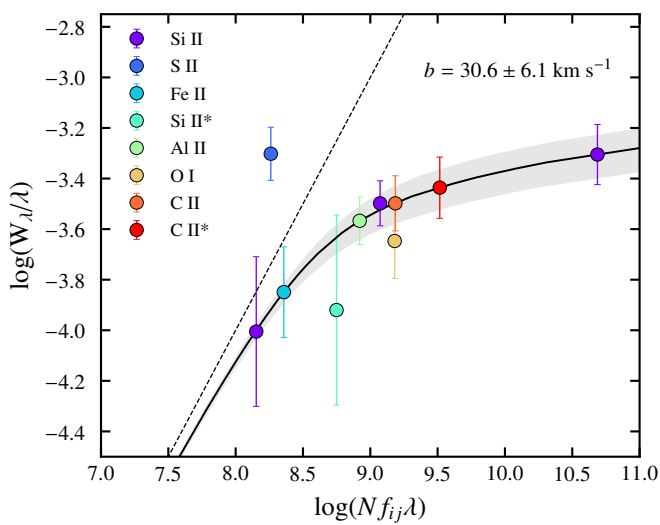

**Fig. 8.** The results obtained with the CoG analysis on the ISM ($z = 3.084$) absorption lines identified in the spectrum of GRB 220627A. The dashed line represents the linear approximation regime ($W_\lambda \propto N$) of the CoG.

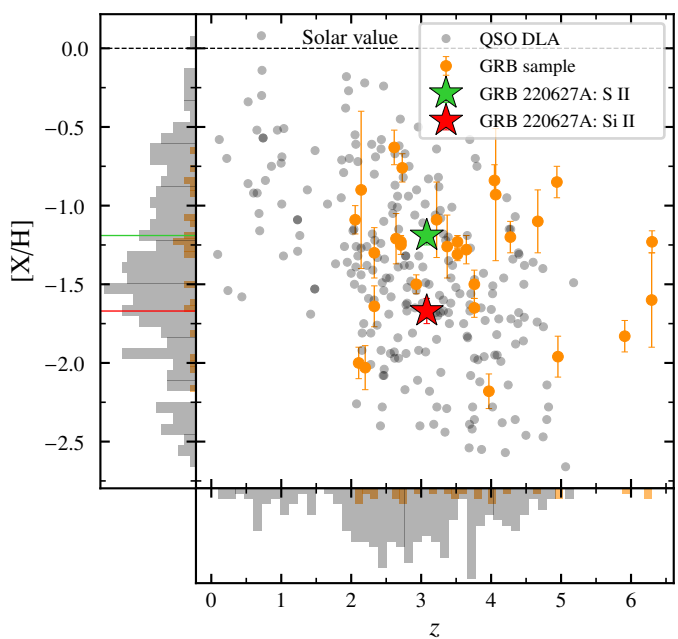

**Fig. 9.** Metallicities obtained from the analysis of S II and Si II absorption lines in GRB afterglows (orange points; Thöne et al. 2013; Bolmer et al. 2019; Saccardi et al. 2023) compared with the corresponding [X/H] values obtained from the same elements in GRB 220627A (green and red data), as a function of redshift. We also show the absorption-derived metallicity values for a large sample of QSO-DLA systems (grey points; Rafelski et al. 2012; De Cia et al. 2016; Saccardi et al. 2023).

In order to reduce the number of free parameters when fitting, we fix the burst redshift at the MUSE value of $z = 3.08$ (Sect. 2.5) and the luminosity distance at $8.25 \times 10^{28}$ cm, computed using our adopted cosmology (Sect. 1). The fraction of shock-accelerated electrons $\xi_N$ is degenerate with respect to the parameters $\{E_{K,iso}, n_0, \epsilon_e, \epsilon_B\}$ (Eichler & Waxman 2005), so we make the common assumption that $\xi_N = 1$. We further assume that the observer is looking directly down the axis of the jet ($\theta_{obs} = 0$). We account for Galactic extinction using the Milky Way extinction curve from Fitzpatrick (1999) with $R_V = 3.1$ and $A_V = 0.13$ mag for the GRB line of sight (Schlafly & Finkbeiner 2011). The host galaxy extinction is a free parameter in our modelling, where we employ the Small Magellanic Cloud (SMC) extinction curve from Pei (1992) at the burst redshift since the vast majority of GRB host galaxies are consistent with an SMC extinction law (Schady et al. 2010; Zafar et al. 2011). We assume that the flux blueward of the rest-frame Lyman limit (912 Å) is totally absorbed. For both Galactic and host extinction we employ the astropy-affiliated `dust_extinction`[6] Python package to redden our model SEDs.

In total we have seven free parameters denoted by the vector $\Theta$:

$$\Theta \equiv \{p, E_{K,iso}, n_0, \epsilon_e, \epsilon_B, \theta_j, A_{V,host}\}. \quad (1)$$

Given a data set $D$ of flux measurements at frequencies $\nu_i$ and times $t_i$, the posterior probability $p(\Theta|D)$ is proportional to the product of the likelihood $p(D|\Theta)$ and the prior $p(\Theta)$ via Bayes'

---
[6] https://dust-extinction.readthedocs.io/en/stable/

Theorem

$$p(\Theta|D) \propto p(D|\Theta)p(\Theta). \quad (2)$$

To account for both detections and non-detections in our data, we employ the likelihood function for Gaussian errors used by Laskar et al. (2014)

$$p(D|\Theta) = \prod p(e_i)^{\delta_i} F(e_i)^{1-\delta_i}, \quad (3)$$

where $\delta_i$ is equal to one for a detection and zero for an upper limit, $e_i$ are the residuals

$$e_i = F_i(\nu_i, t_i) - F_{\text{model}}(\nu_i, t_i; \Theta), \quad (4)$$

$p(e_i)$ is the probability density function of the residuals, and $F(e_i)$ is the cumulative distribution function of the residuals. The two functions $p(e_i)$ and $F(e_i)$ take the form

$$p(e_i) = \frac{1}{\sqrt{2\pi}\sigma_i} e^{-e_i^2/2\sigma_i^2} \quad (5)$$

and

$$F(e_i) = \frac{1}{2}\left[1 + erf\left(\frac{e_i}{\sqrt{2}\sigma_i}\right)\right] \quad (6)$$

respectively, where $\sigma_i$ are the errors on the flux measurements or upper limits. As in Laskar et al. (2014), we also enforce a floor of 5% on the measured flux errors so as not to give undue weight to very high S/N measurements (usually those at optical wavelengths).

`ScaleFit` requires certain parameters to be transformed into dimensionless values in log space in order to improve performance when fitting. The energy $E_{K,\text{iso}}$ is transformed into dimensionless units of $\log 10^{53}$ erg, and the quantity $\log \bar{\epsilon}_e$ is used by `ScaleFit` rather than $\log \epsilon_e$. The two quantities are related via

$$\bar{\epsilon}_e = \frac{p-2}{p-1}\epsilon_e. \quad (7)$$

The circumburst medium density $n_0$ is defined as the number of particles per cm$^3$ at the reference distance of $10^{17}$ cm from the GRB central engine. In a stellar wind medium the circumburst medium has a density profile following $\rho(r) = Ar^{-2}$. Chevalier & Li (2000) define a dimensionless $A_\star$ parameter according to $A_\star = A/(5 \times 10^{11} \text{ g cm}^{-1})$. The $n_0$ parameter is therefore related to $A_\star$ via

$$n_0 = A_\star \times 29.89 \text{ cm}^{-3}. \quad (8)$$

We employ wide, uninformative log-uniform priors on all parameters except for $p$ and $A_{V,\text{host}}$, for which we use uniform priors (see Table 3). The jet opening angle is restricted to a range varying from a very narrow jet ($\theta_j = 0.6°$) to a spherical jet ($\theta_j = 90°$). Our priors are equivalent to those used by Aksulu et al. (2022). During our `emcee` run we utilise 1000 walkers performing 2000 steps through the parameter space. The initial state of the walkers are clustered around the values from a Maximum-Likelihood fit performed using the `scipy` package. We discard the initial 200 steps as burn-in. We performed the MCMC analysis in both an ISM and stellar wind circumburst medium and calculated log likelihoods of $\ln P(\Theta_{\text{ISM}}|D) = 124.9^{+1.6}_{-2.2}$ and $\ln P(\Theta_{\text{Wind}}|D) = 120.0^{+2.5}_{-3.8}$ for each medium, respectively. Assuming both an ISM and stellar wind medium have equal prior probabilities of fitting our data, the Bayes factor is simply the likelihood ratio between the two models. We therefore derive a Bayes factor of 134.29 in favour of an ISM environment. According to Kass & Raftery (1995) a Bayes factor > 100 provides decisive evidence in favour of one model over another, so we only show the results of our ISM modelling henceforth. For completeness we present the results of the wind modelling in Appendix A.

**Table 3.** Priors on parameters in $\Theta$ used in our MCMC analysis.

| Parameter | Prior | Distribution |
|---|---|---|
| $p$ | [1.0, 3.0] | uniform |
| $\log E_{K,\text{iso},53}$ | [−3, 3] | log-uniform |
| $\log n_0$ | [−3, 3] | log-uniform |
| $\log \bar{\epsilon}_e$ | [−10, 1] | log-uniform |
| $\log \epsilon_B$ | [−10, 0] | log-uniform |
| $\log \theta_j$ | [−2, 0.2] | log-uniform |
| $A_{V,\text{host}}$ | [0, 10] | uniform |

### 4.2. Results

Figure 10 shows the marginalised distributions and correlations between pairs of parameters, and Table 4 presents the median values along with their $1\sigma$ uncertainties derived from the marginalised distributions. For the electron spectral index we derive a value of $p = 2.00^{+0.08}_{-0.06}$. This value is consistent within uncertainties with the value for $p$ derived from the post-jet break decay index of $p = \alpha_2 = 2.17 \pm 0.14$ in Sect. 3.2. Furthermore, the model associated with the median values for each parameter (bold lines in Fig. 11) has $\nu_c$ below the optical and X-ray regimes during our observations, consistent with the argument in Sect. 3.3 based on the optical spectral index. This model has $\nu_a \approx 10^{12}$ Hz until $\nu_m$ crosses $\nu_a$ at 0.13 days, whereafter the afterglow spectral breaks are ordered $\nu_m < \nu_a < \nu_c$. The single radio detection at 17 GHz and MeerKAT non-detections at 1.4 GHz are unable to constrain the location of $\nu_m$ and $\nu_a$ precisely, and as a result there are strong degeneracies between $E_{K,\text{iso}}$, $n_0$, $\bar{\epsilon}_e$ and $\epsilon_B$, as expected. In contrast, the jet-opening angle is well-constrained to $4.5^{+1.2}_{-0.3}$ deg and is driven by the clear break in the optical light curves as discussed in Sect. 3.2. The corresponding beaming correction of $f_b = (1 - \cos \theta_j) = 3.1^{+1.8}_{-0.3} \times 10^{-3}$ results in a beaming-corrected prompt $\gamma$-ray energy of $E_\gamma = 1.5^{+0.9}_{-0.2} \times 10^{52}$ erg and kinetic energy of $E_K = 3.1^{+20.3}_{-2.5} \times 10^{51}$ erg. The radiative efficiency of a GRB is defined as (Lloyd-Ronning & Zhang 2004)

$$\eta_\gamma = \frac{E_\gamma}{E_\gamma + E_K}, \quad (9)$$

from which we derive a high radiative efficiency of $\eta_\gamma \approx 84\%$. As mentioned in Sect. 4.1, assuming $\xi_N = 1$ means that our derived kinetic energy is actually a lower limit on the true energy, and therefore this calculated radiative efficiency is actually an upper limit. We note also that the uncertainty on $E_K$ is rather large due to degeneracies in our theoretical modelling. Taking into account the uncertainties in $E_K$ and $E_\gamma$, we derive a $1\sigma$ range of $39-97\%$ for $\eta_\gamma$. Wang et al. (2015) find that radiative efficiencies peak at ~6% with the $1\sigma$ range extending up to ~40%, though they still have a very wide distribution extending to almost 90%, fully consistent with our derived value.

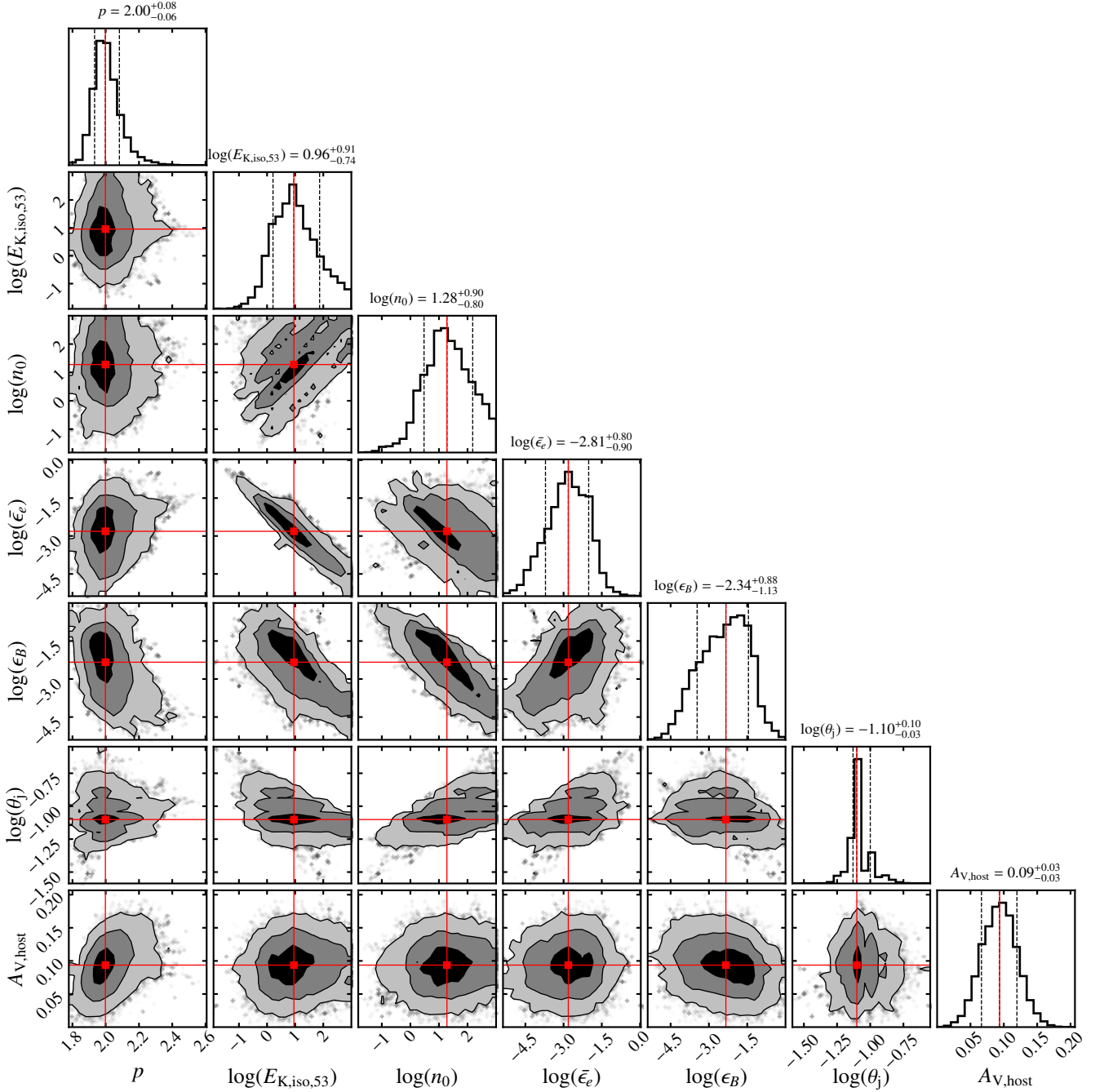

**Fig. 10.** Corner plot from our MCMC implementation showing the 2-D marginalised distributions between pairs of model parameters along with the marginalised distributions for each individual model parameter. Contours denote the $1\sigma$, $2\sigma$ and $3\sigma$ levels while the red lines denote the median values from the marginalised distributions for each model parameter. Uncertainties on each parameter are at the 16$^{\text{th}}$ and 84$^{\text{th}}$ ($1\sigma$) percentile.

## 5. Discussion

The defining feature of GRB 220627A is its two distinct prompt $\gamma$-ray emission episodes separated by ∼600 s, resulting in a total burst duration of ∼1090 s. At a redshift of $z = 3.08$, GRB 220627A is the most distant ultra-long GRB detected to date. We now discuss whether the prompt and afterglow emission are typical for normal long GRBs.

### 5.1. Prompt emission properties

Based on the sample of long GRBs fitted with cutoff power laws in the Fermi-GBM Gamma-Ray Burst Spectral Catalog, Poolakkil et al. (2021) find a distribution of low-energy photon indices of $\Gamma = 1.01^{+0.39}_{-0.35}$ and peak energies of $E_p = 205^{+374}_{-109}$ keV. The prompt emission spectral parameters of GRB 220627A (see Sect. 2.2) are fully consistent with these distributions. For the first emission episode, an additional power law spectral component extending up to higher energies (>100 MeV) is needed to explain the LAT emission. Such a feature has been observed in several other long GRBs (e.g. GRBs 090902B, 090926A;

**Table 4.** Parameter values derived from our theoretical modelling.

| Parameter | value |
| --- | --- |
| $p$ | $2.00^{+0.08}_{-0.06}$ |
| $E_{K,iso}$ ($10^{53}$ erg) | $9.0^{+65.1}_{-7.4}$ |
| $n_0$ (cm$^{-3}$) | $1.9^{+13.3}_{-1.6} \times 10^1$ |
| $\bar{\epsilon}_e$ | $1.5^{+8.1}_{-1.3} \times 10^{-3}$ |
| $\epsilon_B$ | $4.6^{+30.7}_{-4.3} \times 10^{-3}$ |
| $\theta_j$ (deg) | $4.5^{+1.2}_{-0.3}$ |
| $A_{V,host}$ (mag) | $0.09 \pm 0.03$ |

**Notes.** We report the median values and 68% confidence intervals derived from our marginalised distributions in Fig. 10.

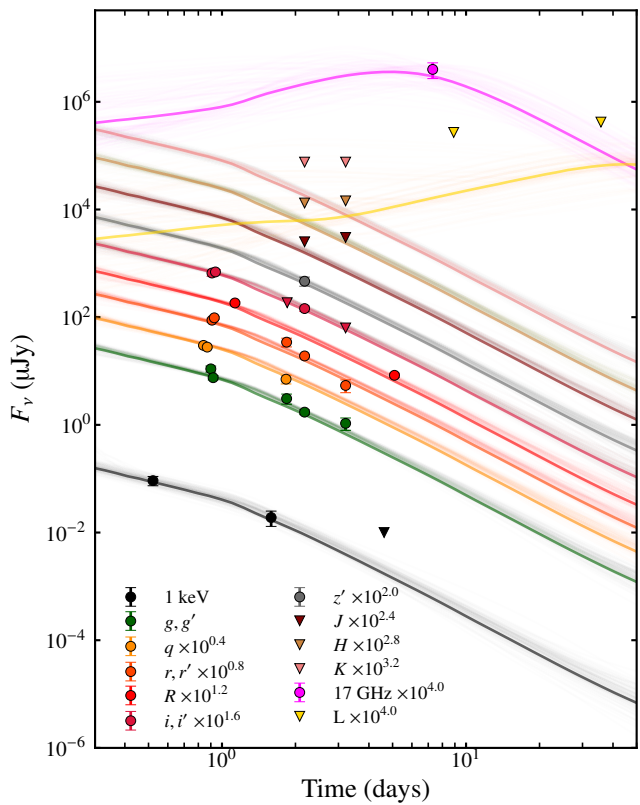

**Fig. 11.** We show all afterglow flux measurements presented in Table 1 along with 100 random models drawn from our MCMC exploration of the parameter space. The bold lines represent the model associated with the median values from the marginalised posterior distributions in Fig. 10. Upper limits are shown as upside-down triangles.

Abdo et al. 2009; Ackermann et al. 2011). The spectral properties of GRB 220627A are therefore not unusual compared to the long GRB population. The temporal features, however, are rare though not unprecedented, and we discuss these further in Sect. 5.3 below.

### 5.2. Afterglow properties

Most of our model parameters derived from our theoretical modelling (Sect. 4.2) are typical values within the literature. Our value of $p$ is within the $1\sigma$ range found by Wang et al. (2015) of $p = 2.33 \pm 0.48$. Wang et al. (2015) also find that the typical isotropic kinetic energy for GRBs without energy injection is $\log(E_{K,iso}/\text{erg}) = 54.66 \pm 1.18$. Our inferred kinetic energy of $E_{K,iso} = 9.0 \times 10^{53}$ erg is well within their sample range. Our circumburst medium density of 19 cm$^{-3}$ agrees within the range of $0.06 - 26$ cm$^{-3}$ found by Aksulu et al. (2022) for long GRBs in ISM environments. Santana et al. (2014) found that $\epsilon_B$ spans almost five orders of magnitude from $10^{-5} - 10^{-1}$. Our value of $\epsilon_B \approx 5 \times 10^{-3}$ is close to their median value of $1.0 \times 10^{-2}$. We are unable to determine a physical value for $\epsilon_e$ using Eq. 7 since $p = 2$ results in $\bar{\epsilon}_e = 0$. Values of $p$ less than two have been suggested in the literature (Panaitescu & Kumar 2002; Dai & Cheng 2001) and have been found through modelling of actual GRB datasets (Aksulu et al. 2022). As discussed in Granot & Sari (2002), using $\bar{\epsilon}_e$ is still valid when $p < 2$ as long as $\gamma_m$ is proportional to the shock Lorentz factor; a high-energy cutoff in the electron energy distribution is then needed so that the total energy in electrons is bounded. Wang et al. (2015) found that the jet opening angle takes values of $\theta_j = 2.5 \pm 1.5$ deg. Our value of $4.5^{+1.2}_{-0.3}$ deg is just outside their $1\sigma$ range, but well within the range of the sample considered by Laskar et al. (2014) having $\theta_j = 7.4^{+11}_{-6.6}$ deg.

The afterglow properties derived from our theoretical modelling are fully consistent with long GRBs, and do not suggest a different or unusual progenitor. The large distance to GRB 220627A means that we are also unable to shed light on the progenitor through observations of a possible supernova (e.g. SN 2011kl accompanying the ultra-long GRB 111209A; Greiner et al. 2015) or another thermal transient. The findings presented here cannot exclude that GRB 220627A has a different progenitor compared to normal long GRBs.

### 5.3. GRBs with widely-spaced emission episodes

Among ultra-long GRBs, GRB 220627A is one of an even smaller subset of GRBs with distinct $\gamma$-ray emission episodes separated by long quiescent periods. Virgili et al. (2013) compared the ultra-long GRB 091024A with a sample of 11 other bursts having durations > 1000 s and divided ultra-long GRBs into those with continuous prompt $\gamma$-ray emission, and those with distinct emission episodes separated by long quiescent periods (so-called interrupted bursts). They pointed out that the total duration of interrupted bursts holds less significance than for continuous bursts, since long quiescent intervals between sub-bursts may not accurately reflect the duration of central engine activity. They identified four bursts with interrupted emission: GRBs 020410A, 080407A, 091024A, and 110709B. Since then, three GRBs have triggered Fermi/GBM twice as listed in the Fourth Fermi/GBM Gamma-Ray Burst Catalog (von Kienlin et al. 2020): GRB 130925A, GRB 150201A, and GRB 160625B. Although GRBs 110709B and 160625B are not strictly ultra-long due to having durations <1000 s, we still include them in this sample since they are close to 1000 s and have widely-spaced emission episodes. Some properties of these bursts including GRB 220627A are presented in Table 5, similar to Table 6 in Virgili et al. (2013). We now discuss the prompt and afterglow properties of these bursts in order to determine if they share any unifying characteristics.

The prompt emission light curves for all eight bursts are diverse in their morphology. Some bursts have weak precursor-like episodes preceding the main episode which contains the majority of the total fluence (GRBs 091024A, 130925A, 160625B), while others have brighter initial episodes (020410A, 080407A, 150201A, 220627A). GRB 110709B has two emission episodes of similar brightness. All bursts appear to show hard to soft evolution across the whole duration of $\gamma$-ray activity, a common trait for the wider GRB population. GRB 220627A is the most distant of all these bursts, and has the highest isotropic γ-ray energy. The bursts in Table 5 span almost two orders of magnitude in $E_{\gamma,\text{iso}}$, comparable to the range of energies of the normal long GRB population (see Fig. 19 in Poolakkil et al. 2021). GRB 220627A is also one of only two bursts that has been detected at GeV γ-ray energies by *Fermi*/LAT.

The prompt emission light curve of GRB 220627A is most similar to that of GRB 110709B, a so-called double burst which triggered *Swift*/BAT twice separated by an interval of 11 minutes. Zhang et al. (2012) studied the prompt and X-ray emission in detail and ruled out a gravitational lensing and a giant X-ray flare origin for the second sub-burst. Their preferred physical explanation for the two sub-bursts with a long quiescent period in between was a magnetar to black hole (BH) scenario, where the first sub-burst is produced by a newly-formed magnetar releasing rotational energy via a magnetised jet, and the second, softer sub-burst is produced by a slower jet once the magnetar has collapsed to a BH. GRB 160625B was an extremely bright burst with three distinct emission episodes which were studied in great detail by Zhang et al. (2018). The first, short precursor episode was thermal-dominated, while the latter two episodes were consistent with synchrotron emission. This spectral transition was interpreted as direct evidence of the GRB jet changing its composition, with the first thermal component associated with the initial core collapse and break out of the ejecta from the surface of the Wolf-Rayet progenitor. For a black hole central engine the quiescent phase prior to the main emission was suggested to be a result of a magnetic barrier delaying accretion until enough material has built up (Proga & Zhang 2006), while for the magnetar engine the delay may be a result of the hot proto-magnetar taking time to dissipate energy before launching an outflow (Komissarov & Barkov 2007; Metzger et al. 2011). Thereafter, a Poynting-flux dominated outflow characterised by synchrotron emission would be launched. Due to the weakness and short duration ($T_{90}$~0.84 s) of the thermal precursor emission, the authors noted that such a feature would not have been observable had GRB 160625B been more distant. Such a feature may have been observed in GRB 220627A had it been closer. Based on 3D simulations, Gottlieb et al. (2022) proposed that the observed quiescent periods of ~1−100 s in prompt emission light curves are naturally explained by a time-varying tilt of the central engine, such that quiescent periods correspond to times when the jet axis is pointed away from the observer. The much longer quiescent period of ~600 s for GRB 220627A, however, may exclude this scenario.

In terms of afterglow emission, almost all of the bursts in Table 5 were detected in X-rays, and their X-ray luminosities are comparable to the majority of the GRB population (see Fig. 12). Levan et al. (2014) regarded the presence of X-ray flares as a defining feature of ultra-long GRBs. GRBs 110709B and 130925A show evidence for X-ray flares, though in the cases of GRBs 091024A and 160625B *Swift*/XRT was unable to observe at early times, so any flares may simply have been undetected (Fig. 12). GRB 150201A does not show any X-ray flares, despite there being early-time data. GRB 220627A was not localised until 0.84 days, so no search for flaring could be conducted. GRB 130925A is the nearest of the ultra-long GRBs in the sample and also has the longest duration measured in γ-rays. Its X-ray afterglow showed many large flares, similar to those of the ultra-long GRBs studied by Levan et al. (2014). Evidence for a thermal emission component was found in the X-ray spectra and was interpreted as photospheric emission from the GRB fireball (Piro et al. 2014). Such thermal emission has been observed in a number of other X-ray flares (Peng et al. 2014). Evans et al. (2014) suggested that the lack of external shock X-ray emission implied a low circumburst density, which could explain the ultra-long γ-ray duration since the deceleration time and radius would be larger, allowing more ejected shells to interact and produce prompt emission before reaching the external shock. The radio emission from GRB 130925A proved challenging to explain in a normal forward shock afterglow model (Horesh et al. 2015). GRBs 091024A, 130925A, and 160625B were all detected and localised by *Swift*/BAT and were therefore monitored from very early times at optical frequencies.

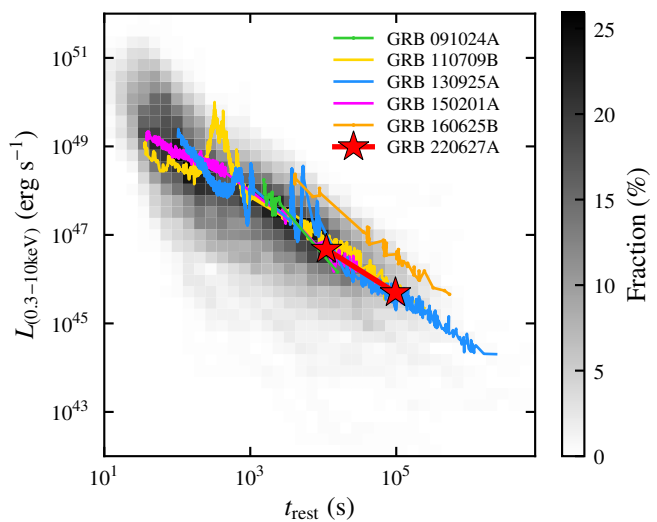

**Fig. 12.** X-ray luminosities as a function of rest-frame time for all *Swift*/XRT-detected bursts in Table 5. For GRBs without redshifts (110709B and 150201A) we assume $z = 1$. The grey colour bar denotes the fraction of *Swift*/XRT-detected afterglows with known redshifts crossing each luminosity-time bin.

In summary, GRBs with widely-spaced emission episodes show a wide variety of prompt and afterglow features. More observations of such bursts are needed in order to determine if there are unifying characteristics that distinguish them from single-emission episode bursts.

## 6. Conclusion

GRB 220627A is an ultra-long GRB that triggered *Fermi*/GBM twice separated by a 956 s time interval. High-energy GeV γ-rays were detected by *Fermi*/LAT in coincidence with the first emission episode which led to the discovery of the optical afterglow with MeerLICHT at 0.84 days. Optical spectroscopic observations with MUSE were used to deduce the burst redshift of $z = 3.08$, making GRB 220627A the most distant ultra-long GRB observed to date. From our optical follow-up observations we identified a temporal break in the light curve at ~1.2 days which we interpret as a jet break. From our spectroscopic measurements we found that the environment of the burst is consistent with a sub-solar metallicity typical of DLA systems and other GRB afterglows. Combined with publicly-available X-ray and radio data, we performed broad-band theoretical modelling of the afterglow data and found preference for a homogeneous circumburst medium. Our most well-constrained parameter is the jet opening angle, which we constrain to $\theta_j = 4.5^{+1.2}_{-0.3}$ deg. All our parameters are typical for long GRB afterglows. Our observations of this ultra-long GRB do not require a different pro-

**Table 5.** Ultra-long GRBs with widely-spaced emission episodes.

| GRB | Duration[a] (s) | $z$ | Fluence[b] ($10^{-5}$ erg cm$^{-2}$) | Energy range[c] (keV) | $E_{\gamma,\mathrm{iso}}$[d] ($10^{52}$ erg) | Episodes[e] | GeV[f] | Afterglow[g] | Refs. |
|---|---|---|---|---|---|---|---|---|---|
| 020410A | 1550 | ~0.5 | 2.8 | 15–1000 | ~1.8 | 4 | - | X,O | 1,2 |
| 080407A | 2100 | - | 44 | 20–1000 | ~120 | 2 | - | - | 3 |
| 091024A | 1300 | 1.092 | 1.5 | 10–10000 | 44 | 3 | ND | X,O | 4,5 |
| 110709B | 900 | - | 0.22 | 15–150 | ~10 | 2 | - | X,R | 6 |
| 130925A | 4500 | 0.347 | 62 | 20–10000 | 19 | 3 | ND | X,O,R | 7,8,9,10,11 |
| 150201A[h] | 1400 | - | 6.8 | 10–1000 | ~36 | 2 | - | X | 12 |
| 160625B | 800 | 1.406 | 66 | 10–1000 | 334 | 3 | D | X,O,R | 13,14 |
| **220627A** | 1200 | 3.08 | 4.5 | 10–1000 | 480 | 2 | D | X,O,R | 15, this work |

**References.** (1) Nicastro et al. (2004); (2) Levan et al. (2005); (3) Pal'shin et al. (2012); (4) Gruber et al. (2011); (5) Virgili et al. (2013); (6) Zhang et al. (2012); (7) Golenetskii et al. (2013); (8) Evans et al. (2014); (9) Piro et al. (2014); (10) Greiner et al. (2014); (11) Horesh et al. (2015); (12) Yu & Pelassa (2015); (13) Zhang et al. (2018); (14) Alexander et al. (2017); (15) Huang et al. (2022).

**Notes.**
[a] Approximate duration of γ-ray activity across all emission episodes.
[b] Fluence as measured across all emission episodes.
[c] Energy range over which reported fluence was measured.
[d] We assume $z = 1$ to calculate $E_{\gamma,\mathrm{iso}}$ for bursts without measured redshifts.
[e] Number of widely-spaced emission episodes present in prompt light curve.
[f] GeV γ-ray emission detected with *Fermi*/LAT. ND indicates a non detection, D indicates a detection.
[g] Afterglow detection in X-rays (X), optical (O) and radio (R).
[h] GRB 150201A triggered *Fermi*/GBM twice separated by 23 minutes. The burst position was earth-occulted during the second trigger, so the association between the two triggers is not beyond doubt.

genitor compared to normal long GRBs as suggested by some authors, though we cannot exclude this possibility. More observations of bursts with widely-spaced emission episodes are needed to determine if they form a separate population with distinct prompt and afterglow features.

## Acknowledgements

The MeerLICHT consortium is a partnership between Radboud University, the University of Cape Town, the South African Astronomical Observatory (SAAO), the University of Oxford, the University of Manchester and the University of Amsterdam, in association with and, partly supported by, the South African Radio Astronomy Observatory (SARAO), the European Research Council and The Netherlands Research School for Astronomy (NOVA). We acknowledge the use of the Inter-University Institute for Data Intensive Astronomy (IDIA) data intensive research cloud for data processing. IDIA is a South African university partnership involving the University of Cape Town, the University of Pretoria and the University of the Western Cape. S.dW. and P.J.G. are supported by NRF SARChI Grant 111692. The research leading to these results has received funding from the European Union's Horizon 2020 Programme under the AHEAD2020 project (grant agreement number 871158). Part of the funding for GROND (both hardware and personnel) was generously granted from the Leibniz-Prize to G. Hasinger (DFG grant HA 1850/28-1) and by the Thüringer Landessternwarte Tautenburg. This work made use of data supplied by the UK Swift Science Data Centre at the University of Leicester.

## References

Abdo, A. A., Ackermann, M., Ajello, M., et al. 2009, ApJ, 706, L138
Ackermann, M., Ajello, M., Asano, K., et al. 2011, ApJ, 729, 114
Aksulu, M. D., Wijers, R. A. M. J., van Eerten, H. J., & van der Horst, A. J. 2020, MNRAS, 497, 4672
Aksulu, M. D., Wijers, R. A. M. J., van Eerten, H. J., & van der Horst, A. J. 2022, MNRAS, 511, 2848
Alexander, K. D., Laskar, T., Berger, E., et al. 2017, ApJ, 848, 69
Aptekar, R. L., Frederiks, D. D., Golenetskii, S. V., et al. 1995, Space Sci. Rev., 71, 265
Asplund, M., Grevesse, N., Sauval, A. J., & Scott, P. 2009, ARA&A, 47, 481
Atwood, W. B., Abdo, A. A., Ackermann, M., et al. 2009, ApJ, 697, 1071
Bacon, R., Accardo, M., Adjali, L., et al. 2010, in Society of Photo-Optical Instrumentation Engineers (SPIE) Conference Series, Vol. 7735, Ground-based and Airborne Instrumentation for Astronomy III, ed. I. S. McLean, S. K. Ramsay, & H. Takami, 773508
Blandford, R. D. & McKee, C. F. 1976, Physics of Fluids, 19, 1130
Bloemen, S., Groot, P., Woudt, P., et al. 2016, in Society of Photo-Optical Instrumentation Engineers (SPIE) Conference Series, Vol. 9906, Ground-based and Airborne Telescopes VI, ed. H. J. Hall, R. Gilmozzi, & H. K. Marshall, 990664
Bolmer, J., Ledoux, C., Wiseman, P., et al. 2019, A&A, 623, A43
Bromberg, O., Nakar, E., Piran, T., & Sari, R. 2013, ApJ, 764, 179
Burrows, D. N., Hill, J. E., Nousek, J. A., et al. 2005, Space Sci. Rev., 120, 165
Chambers, K. C., Magnier, E. A., Metcalfe, N., et al. 2016, arXiv e-prints, arXiv:1612.05560
Chevalier, R. A. & Li, Z.-Y. 2000, ApJ, 536, 195
Dai, Z. G. & Cheng, K. S. 2001, ApJ, 558, L109
De Cia, A., Ledoux, C., Mattsson, L., et al. 2016, A&A, 596, A97
de Wet, S., Groot, P. J., Malesani, D. B., et al. 2022, GRB Coordinates Network, 32289, 1
D'Elia, V., Campana, S., Covino, S., et al. 2011, MNRAS, 418, 680
D'Elia, V., Fiore, F., Meurs, E. J. A., et al. 2007, A&A, 467, 629
di Lalla, N., Axelsson, M., Arimoto, M., Omodei, N., & Crnogoreeviae, M. 2022, GRB Coordinates Network, 32283, 1
Drlica-Wagner, A., Ferguson, P. S., Adamów, M., et al. 2022, ApJS, 261, 38
Eichler, D. & Waxman, E. 2005, ApJ, 627, 861
Evans, P. A., Beardmore, A. P., Page, K. L., et al. 2009, MNRAS, 397, 1177
Evans, P. A., Beardmore, A. P., Page, K. L., et al. 2007, A&A, 469, 379
Evans, P. A. & Swift Team. 2022, GRB Coordinates Network, 32284, 1
Evans, P. A., Willingale, R., Osborne, J. P., et al. 2014, MNRAS, 444, 250
Fitzpatrick, E. L. 1999, PASP, 111, 63
Foreman-Mackey, D., Hogg, D. W., Lang, D., & Goodman, J. 2013, PASP, 125, 306
Frederiks, D., Lysenko, A., Ridnaya, A., et al. 2022, GRB Coordinates Network, 32295, 1
Fynbo, J. P. U., Watson, D., Thöne, C. C., et al. 2006, Nature, 444, 1047
Gao, H., Lei, W.-H., Zou, Y.-C., Wu, X.-F., & Zhang, B. 2013, New A Rev., 57, 141
Gendre, B., Stratta, G., Atteia, J. L., et al. 2013, ApJ, 766, 30
Giarratana, S., Leung, J., Wang, Z., et al. 2022, GRB Coordinates Network, 32454, 1

Golenetskii, S., Aptekar, R., Frederiks, D., et al. 2013, GRB Coordinates Network, 15260, 1
Gottlieb, O., Liska, M., Tchekhovskoy, A., et al. 2022, ApJ, 933, L9
Granot, J. & Piran, T. 2012, MNRAS, 421, 570
Granot, J. & Sari, R. 2002, ApJ, 568, 820
Greiner, J., Bornemann, W., Clemens, C., et al. 2008, PASP, 120, 405
Greiner, J., Mazzali, P. A., Kann, D. A., et al. 2015, Nature, 523, 189
Greiner, J., Yu, H. F., Krühler, T., et al. 2014, A&A, 568, A75
Groot, P. J., de Wet, S., Vreeswijk, P. M., & Meerlicht Consortium. 2022, GRB Coordinates Network, 32281, 1
Gruber, D., Krühler, T., Foley, S., et al. 2011, A&A, 528, A15
Heintz, K. E., Watson, D., Jakobsson, P., et al. 2018, MNRAS, 479, 3456
Horesh, A., Cenko, S. B., Perley, D. A., et al. 2015, ApJ, 812, 86
Huang, Y.-Y., Zhang, H.-M., Yan, K., Liu, R.-Y., & Wang, X.-Y. 2022, ApJ, 940, L36
Izzo, L., D'Elia, V., de Ugarte Postigo, A., et al. 2022, GRB Coordinates Network, 32291, 1
Jakobsson, P., Hjorth, J., Fynbo, J. P. U., et al. 2004, A&A, 427, 785
Jenkins, E. B. 2009, ApJ, 700, 1299
Kass, R. E. & Raftery, A. E. 1995, Journal of the American Statistical Association, 90, 773
Komissarov, S. S. & Barkov, M. V. 2007, MNRAS, 382, 1029
Kouveliotou, C., Meegan, C. A., Fishman, G. J., et al. 1993, ApJ, 413, L101
Krogager, J.-K. 2018, arXiv e-prints, arXiv:1803.01187
Laher, R. R., Gorjian, V., Rebull, L. M., et al. 2012, PASP, 124, 737
Laskar, T., Berger, E., Tanvir, N., et al. 2014, ApJ, 781, 1
Leung, J., Wang, Z., An, T., et al. 2022, GRB Coordinates Network, 32341, 1
Levan, A., Nugent, P., Fruchter, A., et al. 2005, ApJ, 624, 880
Levan, A. J., Tanvir, N. R., Starling, R. L. C., et al. 2014, ApJ, 781, 13
Lien, A., Sakamoto, T., Barthelmy, S. D., et al. 2016, ApJ, 829, 7
Lloyd-Ronning, N. M. & Zhang, B. 2004, ApJ, 613, 477
Meegan, C., Lichti, G., Bhat, P. N., et al. 2009, ApJ, 702, 791
Mészáros, P. & Rees, M. J. 1997, ApJ, 476, 232
Mészáros, P. & Rees, M. J. 2001, ApJ, 556, L37
Metzger, B. D., Giannios, D., Thompson, T. A., Bucciantini, N., & Quataert, E. 2011, MNRAS, 413, 2031
Nakauchi, D., Kashiyama, K., Suwa, Y., & Nakamura, T. 2013, ApJ, 778, 67
Nicastro, L., in't Zand, J. J. M., Amati, L., et al. 2004, A&A, 427, 445
Nicuesa Guelbenzu, A., Klose, S., & Rau, A. 2022, GRB Coordinates Network, 32304, 1
Pal'shin, V., Hurley, K., Goldsten, J., et al. 2012, in Gamma-Ray Bursts 2012 Conference (GRB 2012), 40
Panaitescu, A. & Kumar, P. 2002, ApJ, 571, 779
Pei, Y. C. 1992, ApJ, 395, 130
Peng, F.-K., Liang, E.-W., Wang, X.-Y., et al. 2014, ApJ, 795, 155
Péroux, C., Dessauges-Zavadsky, M., D'Odorico, S., Kim, T.-S., & McMahon, R. G. 2007, MNRAS, 382, 177
Piro, L., Troja, E., Gendre, B., et al. 2014, ApJ, 790, L15
Planck Collaboration, Ade, P. A. R., Aghanim, N., et al. 2016, A&A, 594, A13
Poolakkil, S., Preece, R., Fletcher, C., et al. 2021, ApJ, 913, 60
Prochaska, J. X. 2006, ApJ, 650, 272
Prochaska, J. X., Chen, H.-W., Dessauges-Zavadsky, M., & Bloom, J. S. 2007, ApJ, 666, 267
Proga, D. & Zhang, B. 2006, MNRAS, 370, L61
Qin, Y., Liang, E.-W., Liang, Y.-F., et al. 2013, ApJ, 763, 15
Rafelski, M., Wolfe, A. M., Prochaska, J. X., Neeleman, M., & Mendez, A. J. 2012, ApJ, 755, 89
Rastinejad, J. C., Gompertz, B. P., Levan, A. J., et al. 2022, Nature, 612, 223
Rhoads, J. E. 1999, ApJ, 525, 737
Roberts, O. J., Hristov, B., Meegan, C., & Fermi Gamma-ray Burst Monitor Team. 2022, GRB Coordinates Network, 32288, 1
Ryan, G., van Eerten, H., MacFadyen, A., & Zhang, B.-B. 2015, ApJ, 799, 3
Saccardi, A., Vergani, S. D., De Cia, A., et al. 2023, A&A, 671, A84
Santana, R., Barniol Duran, R., & Kumar, P. 2014, ApJ, 785, 29
Sari, R., Piran, T., & Halpern, J. P. 1999, ApJ, 519, L17
Sari, R., Piran, T., & Narayan, R. 1998, ApJ, 497, L17
Savaglio, S., Fall, S. M., & Fiore, F. 2003, ApJ, 585, 638
Savaglio, S., Rau, A., Greiner, J., et al. 2012, MNRAS, 420, 627
Schady, P., Page, M. J., Oates, S. R., et al. 2010, MNRAS, 401, 2773
Schlafly, E. F. & Finkbeiner, D. P. 2011, ApJ, 737, 103
Skrutskie, M. F., Cutri, R. M., Stiening, R., et al. 2006, AJ, 131, 1163
Soto, K. T., Lilly, S. J., Bacon, R., Richard, J., & Conseil, S. 2016, MNRAS, 458, 3210
Spitzer, L. 1998, Physical Processes in the Interstellar Medium
Stetson, P. B. 1987, PASP, 99, 191
Stratta, G., Gendre, B., Atteia, J. L., et al. 2013, ApJ, 779, 66
Thöne, C. C., de Ugarte Postigo, A., Fryer, C. L., et al. 2011, Nature, 480, 72
Thöne, C. C., Fynbo, J. P. U., Goldoni, P., et al. 2013, MNRAS, 428, 3590
Tody, D. 1993, in Astronomical Society of the Pacific Conference Series, Vol. 52, Astronomical Data Analysis Software and Systems II, ed. R. J. Hanisch, R. J. V. Brissenden, & J. Barnes, 173
van Eerten, H., van der Horst, A., & MacFadyen, A. 2012, ApJ, 749, 44
van Eerten, H. J. & MacFadyen, A. I. 2012a, ApJ, 747, L30
van Eerten, H. J. & MacFadyen, A. I. 2012b, ApJ, 751, 155
Virgili, F. J., Mundell, C. G., Pal'shin, V., et al. 2013, ApJ, 778, 54
von Kienlin, A., Meegan, C. A., Paciesas, W. S., et al. 2020, ApJ, 893, 46
Vreeswijk, P. M., Ledoux, C., Smette, A., et al. 2007, A&A, 468, 83
Wang, X.-G., Zhang, B., Liang, E.-W., et al. 2015, ApJS, 219, 9
Wang, X.-G., Zhang, B., Liang, E.-W., et al. 2018, ApJ, 859, 160
Willingale, R., Starling, R. L. C., Beardmore, A. P., Tanvir, N. R., & O'Brien, P. T. 2013, MNRAS, 431, 394
Woosley, S. E. & Bloom, J. S. 2006, ARA&A, 44, 507
Worters, H. L., O'Connor, J. E., Carter, D. B., et al. 2016, in Society of Photo-Optical Instrumentation Engineers (SPIE) Conference Series, Vol. 9908, Ground-based and Airborne Instrumentation for Astronomy VI, ed. C. J. Evans, L. Simard, & H. Takami, 99083Y
Yang, J., Ai, S., Zhang, B.-B., et al. 2022, Nature, 612, 232
Yu, H. F. & Pelassa, V. 2015, GRB Coordinates Network, 17370, 1
Zafar, T., Watson, D., Fynbo, J. P. U., et al. 2011, A&A, 532, A143
Zhang, B. 2018, The Physics of Gamma-Ray Bursts
Zhang, B., Fan, Y. Z., Dyks, J., et al. 2006, ApJ, 642, 354
Zhang, B. & Mészáros, P. 2004, International Journal of Modern Physics A, 19, 2385
Zhang, B.-B., Burrows, D. N., Zhang, B., et al. 2012, ApJ, 748, 132
Zhang, B. B., Zhang, B., Castro-Tirado, A. J., et al. 2018, Nature Astronomy, 2, 69
Zhang, B.-B., Zhang, B., Murase, K., Connaughton, V., & Briggs, M. S. 2014, ApJ, 787, 66
Zhang, W. & MacFadyen, A. 2009, ApJ, 698, 1261

**Table A.1.** Same as Table 4, but for our modelling in a stellar wind medium.

| Parameter | value |
|---|---|
| $p$ | $2.12^{+0.19}_{-0.10}$ |
| $E_{K,iso}$ ($10^{53}$ erg) | $3.7^{+16.1}_{-2.9}$ |
| $n_0$ (cm$^{-3}$) | $1.1^{+3.2}_{-0.8} \times 10^2$ |
| $\bar{\epsilon}_e$ | $4.2^{+16.3}_{-3.1} \times 10^{-3}$ |
| $\epsilon_B$ | $1.9^{+11.8}_{-1.7} \times 10^{-3}$ |
| $\theta_j$ (deg) | $5.7^{+2.9}_{-2.5}$ |
| $A_{V,host}$ (mag) | $0.11 \pm 0.03$ |

## Appendix A: A wind model for GRB 220627A

We performed theoretical modelling as in Sect. 4, but for a stellar wind circumburst medium. The parameter values derived from of our MCMC analysis are presented in Table A.1, and a sample of model light curves are shown in Fig. A.1. Compared to the ISM case, we find that a slightly higher value of $p$ is preferred (2.12 vs 2.00) along with a wider jet (5.7 vs 4.5 deg), though both parameters agree within uncertainties with the ISM values. The wider jet results in a jet break in the light curves at later times than in the ISM case (1.8 vs 1.2 days). The model associated with the median values from the marginalised posterior distributions (bold line in Fig. A.1) undergoes fast cooling ($\nu_c < \nu_m$) until 0.06 days. The rising cooling frequency ($\nu_c \propto t^{1/2}$) is below the optical and X-ray bands until $\sim$10 days, so that the light curves in both observing bands decline as $\alpha = (2-3p)/4 = -1.09$ prior to the jet break at $\sim$1.8 days. The self-absorption frequency is at $\sim10^{12}$ Hz at 0.01 days and declines as $t^{-3/5}$ until $\nu_m$ crosses $\nu_a$ at 0.18 days. All the radio observations are therefore in the self-absorbed regime, and the peak in the light curves are due to the passage of $\nu_m$ through the observing band in question. The jumps in the radio light curves are due to the fact that simulations in a wind environment struggle more than ISM simulations to numerically resolve the radial profile of a relativistic blast wave.

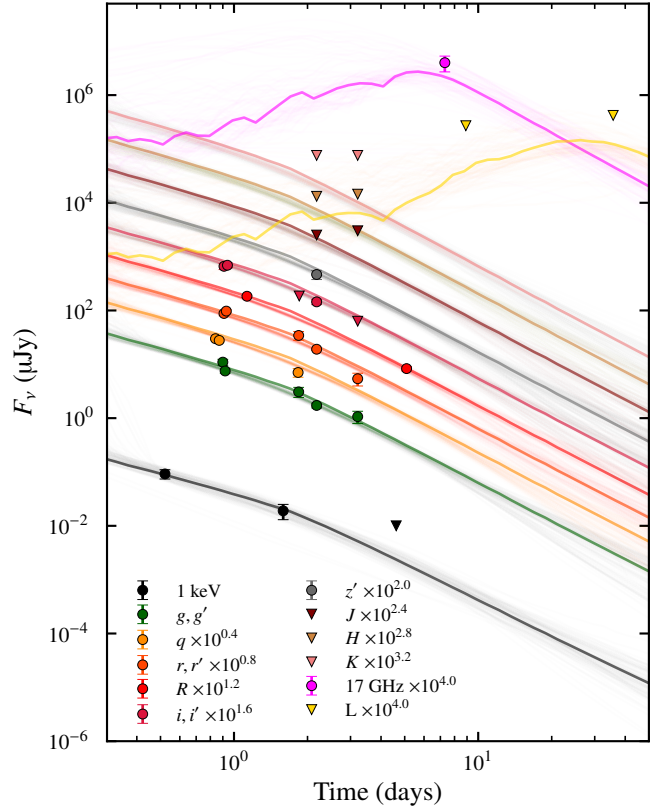

**Fig. A.1.** Same as Fig. 11, but for our modelling in a stellar wind medium.